\documentclass[prd,superscriptaddress,nofootinbib,11pt, tightenlines]{revtex4-2}
\usepackage{multirow}
\usepackage{epsfig}
\usepackage{amsmath}
\usepackage{amsfonts}
\usepackage{xfrac}
\usepackage{float}
\usepackage{mathrsfs}
\usepackage{bm}
\usepackage{amssymb}
\usepackage{slashed}
\usepackage{color}
\usepackage{pbox}
\usepackage{subfigure}
\usepackage{array,multirow,makecell}
\usepackage[colorlinks,citecolor=blue, linkcolor = blue, urlcolor = blue]{hyperref}
\usepackage{tabularx}
\usepackage{scrextend}
\usepackage{natbib}
\usepackage{lipsum}
\usepackage[normalem]{ulem} 
\usepackage{tikz,xcolor,hyperref}
\usepackage{marginnote}
\usepackage{cancel}
\usepackage{subfigure}
\usetikzlibrary{arrows.meta}
\usepackage{tikz}
\usepackage{tikz-feynman}
\tikzfeynmanset{compat=1.1.0}

\newcommand{\beq}{\begin{equation}}
		\newcommand{\eeq}{\end{equation}}

\newcommand{\TeV}{\text{TeV}}

{}
{}

\def \nchi0{\widetilde\chi^0}

\newcommand{\newc}{\newcommand}
\newc{\ba}{\begin{array}}
		\newc{\ea}{\end{array}}
\newc{\bea}{\begin{eqnarray}}
		\newc{\eea}{\end{eqnarray}}
\newc{\beastar}{\begin{eqnarray*}}
		\newc{\eeastar}{\end{eqnarray*}}
\newc{\bestar}{\begin{equation*}}
		\newc{\eestar}{\end{equation*}}
\newc{\ben}{\begin{enumerate}}
		\newc{\een}{\end{enumerate}}

\begin{document}

\title{Radiative Corrections to the Direct Detection of Inelastic Scattering of Higgsino-like Neutralino Dark Matter}

\author{Arindam Chatterjee}
\email{arindam.chatterjee@snu.edu.in}
\affiliation{Shiv Nadar Institution of Eminence deemed to be University, Gautam Buddha Nagar,
	Uttar Pradesh, 201 314, India}

\author{Debottam Das}
\email{debottam@iopb.res.in}
\affiliation{ Institute of Physics, Sachivalaya Marg, Bhubaneswar, 751 005, India}
\affiliation{Homi Bhabha National Institute, Training School Complex, Anushakti Nagar,
	Mumbai 400 094, India}

\author{Syed Adil Pasha}
\email{sp855@snu.edu.in}
\affiliation{Shiv Nadar Institution of Eminence deemed to be University, Gautam Buddha
Nagar, Uttar Pradesh, 201 314, India}
\affiliation{ Institute of Physics, Sachivalaya Marg, Bhubaneswar, 751 005, India}
\affiliation{Homi Bhabha National Institute, Training School Complex, Anushakti Nagar,
	Mumbai 400 094, India}

\author{Alexander Pukhov}
\email{alexander.pukhov@gmail.com}
\affiliation{Skobeltsyn Institute of Nuclear Physics, Moscow State University,
	Moscow, 119 992, Russia}

\author{Rahul Puri}
\email{rahul.puri@iopb.res.in}
\affiliation{ Institute of Physics, Sachivalaya Marg, Bhubaneswar, 751 005, India}
\affiliation{Homi Bhabha National Institute, Training School Complex, Anushakti Nagar,
	Mumbai 400 094, India}

\begin{abstract}
	The direct detection (DD) of Higgsino{-like} dark matter (DM) through inelastic
    scattering processes may provide a promising avenue, along with the elastic
    scattering, when the mass splitting between the neutral Higgsino pairs is extremely
    tiny.
    The mass splitting can be reduced further by adopting the on-shell
    renormalization for the neutral and charged Higgsinos in the MSSM. 
    Moreover, all the one-loop electroweak (EW) corrections to the three-point
    vertices for the neutralino(s)-Higgs interactions have been considered, 
    while both elastic and inelastic DM-nucleon scattering cross-sections have been
    calculated primarily through Higgs exchange. Subsequently, the
    expected number of scattering events in the latest LUX-ZEPLIN (LZ) DD experiment
    are also computed.
    Our results highlight a few scenarios in which the inelastic component may
    exceed the corresponding elastic component, leading to enhanced direct detection
    scattering rates.
\end{abstract}

\maketitle
\section{Introduction}
\label{sec:intro}

Among the low-energy supersymmetric (SUSY) extensions of 
the Standard Model (SM), the Minimal Supersymmetric 
Standard Model (MSSM) with conserved $R$-parity provides 
a well-motivated dark matter (DM) candidate. In large 
regions of the MSSM parameter space, the lightest 
neutralino ($\tilde{\chi}_1^0$) is the lightest 
supersymmetric particle (LSP). Owing to the 
conservation of $R$-parity, it is stable and can 
account for the observed DM relic 
abundance~\cite{Jungman:1995df,Bertone:2004pz}.
In general, $\tilde \chi_1^0$ can be characterised by 
one of the interaction states $-$ Bino, Wino, or 
Higgsino, or by any of their suitable admixtures.
Among various possibilities,  
a nearly pure Higgsino LSP is favoured as it 
offers a potential resolution to the naturalness 
problem \cite{Barbieri:1987fn, Ellis:1986yg, Chan:1997bi, Feng:2013pwa,
Giudice:2013nak, Baer:2012cf, Mustafayev:2014lqa}. 
{The observed dark matter (DM) relic abundance, 
($\Omega_{\rm DM} h^2 \simeq 0.12$)~\cite{WMAP:2012nax,Planck:2018vyg}, can
be naturally accommodated within the Higgsino-like DM 
scenario~\cite{Chan:1997bi,Chattopadhyay:2003xi,Chattopadhyay:2005mv,
Akula:2011jx,Baer:2011ab,Ellis:2012aa,Buchmueller:2013rsa,He:2023lgi}. In
particular, a nearly pure Higgsino-like $\tilde \chi_1^0$ with a mass
parameter ($\mu \simeq 1~\mathrm{TeV}$) yields the correct 
thermal relic density owing to its sizable pair-annihilation cross-sections
into electroweak gauge bosons.}
%

Fine-tuning measures, particularly those defined at 
the electroweak scale, generally favor a Higgsino 
mass parameter in the range $\mu \sim \mathcal{O}
(100$--$1000)\,\text{GeV}$ 
\cite{Baer:2012up,Baer:2012cf}. In such regions of 
parameter space, one typically finds a moderate fine-tuning level,
$\mathcal{O}(10$--$100)$, even when 
multi-TeV third-generation squarks and gluinos are   
{permitted~\cite{Baer:2012up,Baer:2012cf}}.
Supported with a compressed neutralino spectra, Higgsino-like
$\tilde{\chi}_1^0$ can avoid potentially relevant LHC 
{bounds \cite{ATLAS:2025lhc}}. Moreover, such 
scenarios are largely unconstrained by indirect 
searches for DM {annihilation \cite{Cohen:2013ama, Fan:2013faa, MAGIC:2022acl, Dessert:2022evk}}.\footnote{Indirect searches can also probe scenarios
with a Bino-like LSP containing a Higgsino admixture 
\cite{Chattopadhyay:2024qgs}.} For the direct 
detection (DD) of DM, 
tree-level couplings of the Higgs and $Z$ bosons to a 
Higgsino-like $\tilde \chi_1^0$ are suppressed, leading 
to reduced spin-independent and spin-dependent 
scattering cross-sections. The next-to-leading order 
(NLO) corrections {lead} to important effects in 
this specific context, thus have been accounted for, in 
the calculation of the DD cross-sections and DM relic 
{density~\cite{Drees2,Drees1,Drees:1996pk,Hisano:2004pv,Hisano:2011cs, Hisano:2012wm,Baro:2007em, Baro:2009na, Chatterjee:2012hkk, Harz:2014gaa,Harz:2014tma,Klasen:2016qyz, Harz:2023llw}}. In the context of an improved 
calculation of the DD {cross-section of Higgsino-like $\tilde \chi_1^0$}, earlier, Ref.~\cite{Drees:1996pk} evaluated dominant loop 
contributions to the { $\tilde{\chi}_1^0 \tilde{\chi}_1^0 h_j$ vertices ($h_1,h_2$ being the light and heavy Higgs bosons in MSSM, respectively),} arising from
third-generation quark–squark loops, while focusing on relatively light Higgsino-like
neutralinos and incorporating approximate 
effects of gaugino–Higgsino mixing. 

Subsequently, Refs.~\cite{Hisano:2004pv,Hisano:2011cs,
Hisano:2012wm} investigated important radiative 
effects involving electroweak gauge bosons and box 
diagrams for Higgsino-like (and also Wino-like) 
neutralinos in the pure gauge eigenstate limit, while 
neglecting contributions from fermions and 
sfermions due to their assumed heavy masses. 
Ref.~\cite{Harz:2023llw, Klasen:2016qyz} presented the 
SUSY QCD corrections for the DD of $\tilde \chi_1^0$ 
DM.\footnote{{In this context, a class of SUSY QCD 
corrections to the vertices involving the Higgs bosons 
and quarks have been studied in Ref. \cite{Djouadi:2000ck}, and these have been implemented 
in the publicly available code \texttt{micrOMEGAs} 
\cite{Belanger:2008sj}.}} The DM-nucleon 
{scattering} cross-section at one-loop level for a 
generic class of weakly interacting particles was 
presented in 
Ref.~\cite{Cirelli:2005uq, Hisano:2011cs}. Similarly, 
the interaction of gluons with the DM was presented in 
Ref.~\cite{Hisano:2010fy, Hisano:2010ct}. In the recent 
past, a more refined prediction of the DD of Higgsino-like
$\tilde \chi_1^0$ considers all relevant SUSY 
particles in the three-point $\tilde \chi_1^0 \tilde \chi_1^0 h_j$
vertices \cite{Bisal:2023fgb} 
along with quark \textbf{twist-2}, as discussed in {Ref.~\cite{Drees1}}. In addition, the
NLO contribution to the $\tilde \chi_1^0$--$g$ interaction was also considered
following Ref.~\cite{Drees1}, as implemented in \texttt{micrOMEGAs}\cite{Belanger:2008sj}.\footnote{{Similar studies in the 
context of generic $\tilde \chi_1^0$, and for wino-like 
$\tilde \chi_1^0$ may be found in Refs.~\cite{Bisal:2023iip,Bisal:2024iar}. Further,
radiative corrections to the $\tilde \chi_1^0 \tilde \chi_1^0 Z$ vertex, and
implications for the spin-dependent DD have been studied in Ref.
\cite{Bisal:2024ezn}.}} Considering all these relevant electroweak corrections,
the NLO improved $\tilde \chi_1^0$-nucleon cross-section can delineate 
the \textbf{blind spot}\footnote{ For earlier works on blind spots, see Refs.~\cite{Cheung:2012qy, Cheung:2013dua, Huang:2014xua}} scenarios where
the DM DD experiments cannot confirm the presence of DM\footnote{Another example
follows when the SM-like Yukawa couplings of the light quarks can be considered
to deviate from the SM values~\cite{Das:2020ozo}.} \cite{Bisal:2026hpm}.

In this work, we would like to move towards the next 
step to have the theoretically more precise 
{estimation of Higgsino-like} $\tilde \chi^0_1$-nucleon scattering cross-section. As is known, a 
Higgsino-like LSP is associated with a nearly 
degenerate Higgsino-like next-to-lightest SUSY particle 
(NLSP) with mass splitting $\sim \mathcal O(10-100)$ MeV or even smaller,
for very large gaugino mass parameters of $\mathcal O(10)$ TeV. 
For a DM particle with a mass of ($\mathcal{O}(1)\,
\mathrm{TeV}$), the typical kinetic energy in the 
galactic halo is of ($\mathcal{O}(100)\,\mathrm{keV}$), 
assuming a characteristic velocity ($v\sim 10^{-3}$). 
Consequently, transitions to the nearly degenerate 
heavier state can be kinematically accessible, in 
particular, for mass splittings ($\delta \lesssim \mathcal{O}(100)\, \rm{keV}$).
This framework is commonly referred to as the inelastic DM 
scenario~\cite{Hall:1997ah,Tucker-Smith:2001myb}. 
In this case, the total number of events to elastic DM–nucleon
scattering must be supplemented by its 
inelastic counterpart \cite{Tucker-Smith:2001myb, Tucker-Smith:2004mxa}. 
{Therefore, in addition to elastic scattering, the 
spin-independent inelastic scattering of Higgsino-like DM 
must also be taken into account.}

{It is worth mentioning that, very recently, the LUX-ZEPLIN (LZ) experiment reported a nuclear-recoil event at approximately 248 keV, with a local significance of about 3.4$\sigma$ and a global significance of 2.6$\sigma$ \cite{Akerib:2026jyz}. Although further data are required before attributing this event to dark matter, its relatively large recoil energy has motivated recent studies of TeV-scale Higgsino dark matter and inelastic scattering interpretations \cite{Wu:2026nhi, Freese:2026sga, Fan:2026kxx}. In this context, a precise treatment of both the neutralino mass splitting and the relevant elastic and inelastic scattering amplitudes becomes particularly important. The radiative corrections considered in the present work can therefore be directly relevant for assessing such scenarios.}

{In the same spirit as the evaluation of the elastic DM 
direct-detection cross-section~\cite{Bisal:2023fgb,Bisal:2023iip,Bisal:2024iar},} we compute the radiative corrections to the { $\tilde{\chi}_1^0 \tilde{\chi}_i^0 h_j$ vertices, with $i,j \in \{1,2\}$,  relevant for the present analysis.} 
The {on-shell} renormalization of the chargino-neutralino
sector is considered using appropriate 
variants of the on-shell renormalization scheme 
\cite{Fritzsche:2002bi, Baro2009,Chatterjee:2011wc}, 
which subsequently has been used to evaluate the vertex 
counterterm contributions to the $\tilde \chi_1^0 
\tilde \chi_1^0 h_i$ and $\tilde \chi_1^0 \tilde 
\chi_2^0 h_i$ vertices. {As the relative { contribution} of the inelastic scattering is very 
sensitive to the mass splitting of the two lightest 
neutralinos,} we use the on-shell renormalization 
scheme to calculate the one-loop corrected masses for 
these states. {Notably, {in earlier works~\cite{Bisal:2023fgb,Bisal:2023iip,Bisal:2024ezn,Bisal:2024iar}, radiative corrections to} 
only} the $\tilde \chi_1^0 \tilde \chi_1^0 h_j$ vertices were considered and used in the public code \texttt{micrOMEGAs} \cite{Belanger:2010pz,Belanger:2013oya} to compute the spin-independent elastic scattering cross-section. In the present work, the contributions from the radiatively corrected $\tilde{\chi}_1^0 \tilde{\chi}_2^0 h_j$ vertex diagrams have been considered and 
the relevant inelastic scattering cross-sections are evaluated using { our code
following Refs.~\cite{Drees:1993bu, ELLIS1991, ELLIS1993}}
(see Sec. \eqref{sec:Improv}). Finally, {contributions from the radiatively corrected elastic, as well as inelastic processes} are combined to estimate the event rates, and to evaluate the detection prospects for Higgsino-like $\tilde \chi^0_1$ DM in light of upcoming and future experiments.
Our results demonstrate that inelastic scattering can become the dominant contribution in certain regions of the parameter space, underscoring the importance of including inelastic processes in DM-DD studies.

We note here that the relic density constraint is not 
always respected in the analysis; thus, thermal relic
abundance of the LSP may satisfy (i) the observed 
cosmological dark matter abundance, or (ii) falls short of or overshoots the observed DM abundance. However, a few different proposals may help to attune the observed relic density for the Higgsino-like $\tilde{\chi}_1^0$ DM.  This includes (i) new coannihilating SUSY particles \cite{Chakraborti:2015mra}, (ii) Sommerfeld boost factor for heavy neutralino DM of mass $\geq \mathcal{O}(1~\text{TeV})$  \cite{Hisano:2006nn, Beneke:2014hja,Hisano:2004ds, Mohanty:2010es,Beneke:2019qaa}, (iii) non-thermal production mechanisms \cite{Aparicio:2016qqb}, and (iv) the presence of additional DM components, such as axions \cite{Bae:2013hma}. 
In scenarios where $\tilde{\chi}_1^0$ constitutes only 
a fraction of the total dark matter density, the 
corresponding direct detection limits are 
proportionally weakened. In the present work, we do not 
impose the constraint of reproducing the observed 
thermal relic abundance.\footnote{ 
Note that the mass splitting between $\tilde{\chi}_2^0$ and
$\tilde{\chi}_1^0$ is of $\mathcal{O}(10)$~keV, which is much smaller than
the freeze-out temperature ($T_f \sim m_{\tilde{\chi}_1^0}/20$). Since the
calculation of the relic abundance includes all relevant coannihilation
channels, the inclusion or omission of the small decay width of
$\tilde{\chi}_2^0$ does not affect the predicted relic abundance.
}

The remainder of this article is arranged as follows: 
Sec.~\ref{sec:DM DD} discusses the generalities of
DM direct detection, which involves both elastic and 
inelastic DD cross-sections. The radiative corrections 
of masses and $\tilde{\chi}_1^0 \tilde{\chi}_i^0 h_j$
vertices are discussed in Sec.~\ref{sec:Rad Corr}, and 
improvements beyond \texttt{micrOMEGAs} are
discussed in Sec.~\ref{sec:Improv}. The results are 
presented in Sec.~\ref{sec:results} and the article is concluded in Sec.~\ref{sec:Conclusion}.

\section{Dark Matter-Nucleus scattering in Direct detection}
\label{sec:DM DD}

In this section, we discuss the theoretical framework for 
DM-nucleus scattering in DD 
experiments. While the conventional DM DD searches 
include only the elastic scatterings ($\tilde 
\chi_1^0q\to\tilde\chi_1^0q$), the inelastic interaction 
($\tilde \chi_1^0q\to\tilde\chi_2^0q$) can also play a 
crucial role when the mass splitting $\delta$ between the 
LSP and the NLSP is less or comparable to the typical 
kinetic energies of DM at earth ($T^\odot_{\tilde\chi_1^0}$).
 The total number of scattering events $N_E$ observed in the detector with fiducial mass $M_D$ exposed for a time $t_{\rm ex}$ can be evaluated as~\cite{LEWIN199687},
\begin{align}
    N_E=M_Dt_{\rm ex}\int dE_R~\epsilon(E_R)\frac{dR_{SI}}{dE_R}(E_R)\label{eq:nEvents}
\end{align}
where $\epsilon(E_R)$ is the detector efficiency at recoil energy $E_R$. For the recent {LUX-ZEPLIN (LZ)}~\cite{LZCollaboration:2024lux} experiment the exposure is $M_Dt_{\rm ex}=4.2$\,tonne-yrs. The differential {rate of scattering events (in the lab frame)} 
per unit detector mass is given by \cite{LEWIN199687} {
\begin{align}\label{1}
	\frac{dR_{\rm SI}}{dE_R} = N_T \frac{\rho_{\odot}}{m_{\tilde\chi_1^0}} \sum_{i}{\Theta(v_{\rm max}-v_{\rm min}^{i}(E_R))}\int\limits_{v_{\rm min}^{i}(E_R)}^{v_{\rm max}} dv ~vf(v) 
	\frac{d\sigma_{\rm SI}^{i}}{dE_R}(v, E_R)~;\hspace{0.5cm} i\in\{\rm elastic,~inelastic\}
\end{align}
}where the sum (over $i$) runs over elastic and inelastic channels. 
Here, $E_R$ is the recoil energy of the target nucleus, 
$N_T$ is the number of target nuclei per unit detector 
material mass, $\rho_\odot =0.3\,\rm GeV/cm^3 $ is the 
local DM density \cite{Tucker-Smith:2001myb} with DM mass 
$m_{\tilde\chi_1^0}$. Here, $v$ is the speed of the incident 
DM in the lab frame
{with upper limit $v_{\rm max}=v_{\rm esc}+v_e$, $v_e$ being the
relative velocity of earth with respect to the galactic rest frame
and $v_{\rm esc}$ is the escape speed of the Milky Way galaxy. The
Heaviside $\Theta$ function removes the contributions from
arbitrarily large $v_{\rm min}$ at too large or too small $E_R$
\cite{Tucker-Smith:2004mxa}. Similarly,}
$f(v)$ denotes the DM velocity distribution, $v_{\rm min}^i(E_R)$
is the minimum velocity at which the channel $i$ can impart a 
recoil energy $E_R$ to the target nucleus {without exciting it}, and
is given by 
\begin{align}\label{eq:vminInel}
	v_{\rm min}^{i}(E_R)=\frac{1}{\sqrt{2m_AE_R}}
		\left(\frac{E_R m_A}{\mu_A}+\delta_i\right).
\end{align}
In the above equation, $\delta_{\rm elastic} = 0$, while, for inelastic scattering,
$\delta_{\rm inelastic} = \delta = m_{\tilde{\chi}_2^0}- 
m_{\tilde{\chi}_1^0}$ (see \textbf{Appendix 
\eqref{sec:velocities}}), $m_A$ is the mass of the 
target nucleus, $\mu_A$ is the reduced mass of the DM 
and nucleus and $\dfrac{d\sigma^i_{\rm SI}}{dE_R}$  is the 
differential cross-section for DM-nucleus scattering for 
a given channel $i$. The latter can be evaluated as,{
\begin{align}\label{2}
	\frac{d\sigma_{\rm SI}^{i}}{dE_R}
	 & =\frac{m_A}{2v^2} F^2(E_R)\left[\frac{f_p^{i} Z+f_n^{i}(A-Z)}{{f_p^i}}
	\right]^2\frac{\sigma_p^i}{\mu_p^2},
\end{align}
}where $A$ and $Z$ represent the mass number and the 
atomic number of the target nucleus, respectively, 
{ $\mu_p$ stands for the reduced mass of the proton 
and DM, 
$\sigma_p^i$ denotes the proton cross-section} for the $i$-th { channel} at zero momentum transfer, in the elastic limit, \textit{i.e.}, with $\delta_i = 0$, and $F(E_R)$ denotes the nuclear form factor. { In this work, we use the Helm nuclear form factor~\cite{Engel:1991wq} for our calculations.} Finally, {$f_N^{i}~(N\in\{p,n\})$ represents the effective $\tilde\chi_1^0\tilde\chi_i^0\bar NN$ couplings, relevant for elastic ($i=1$) and inelastic ($i=2$) scattering.
} 
{In the present context, 
for Higgs mediated SI processes $f_n^i \simeq f_p^i$.}
{Further, note that} following Ref.~\cite{Tucker-Smith:2001myb}, the DM velocity distribution $f(v)$ is assumed to be a standard Maxwell-Boltzmann distribution with $v_{\rm rms}=v_0\sqrt{{3}/{2}}$, $v_0=220$\,km/s being the rotational speed of the local standard of rest~\cite{Kerr:1986cei} {(see \textbf{Appendix \eqref{sec:velocities}})}. 


The interaction terms in the effective Lagrangian relevant for DM–nucleon scattering can be categorized as
\begin{align}
\mathscr{L}_{\rm eff} \supset \mathscr{L}^{\rm el}_{\rm eff} + \mathscr{L}^{\rm inel}_{\rm eff}~,
\end{align}
where $\mathscr{L}^{\rm el}_{\rm eff}$ ($\mathscr{L}^{\rm inel}_{\rm eff}$) denotes the set of operators responsible for elastic (inelastic) scattering processes. In the following, we discuss the explicit forms of these operators and present the corresponding scattering amplitudes for each case.

\subsection{Elastic neutralino-nucleon spin-independent scattering}
The effective lagrangian for the elastic neutralino-nucleon scattering at the nucleon mass scale is given as,
\begin{align}
	\mathscr{L}_{\rm eff}^{\rm el} \supset\sum_{u,d,s}\mathscr{L}_q^{\rm el} + \mathscr{L}_g^{\rm el}
\end{align}
where $\mathscr{L}_q^{\rm el},\mathscr{L}_g^{\rm el}$ represents effective $\tilde\chi_1^0\tilde\chi_1^0$--$qq$ and $\tilde\chi_1^0\tilde\chi_1^0$--$gg$ vertices, respectively, and are given as~\cite{Drees1, Drees2,Hisano:2011cs},
\begin{align}
	\mathscr{L}_{q}^{\rm el} & =\lambda_qm_q\bar{\tilde{\chi}}^0_1\tilde{\chi}^0_1 \bar{q}q +\frac{g_q^{(1)}}{m_{\tilde{\chi}^0_1}}\bar{\tilde{\chi}}^0_1 i\partial^\mu\gamma^\nu\tilde{\chi}^0_1\mathcal{O}^q_{\mu\nu}+ \frac{g_q^{(2)}}{m^2_{\tilde{\chi}^0_1}}\bar{\tilde{\chi}}^0_1 (i\partial^\mu)(i\partial^\nu)\tilde{\chi}^0_1\mathcal{O}^q_{\mu\nu} \nonumber \\
	\mathscr{L}_{g}^{\rm el} & =\lambda_g\bar{\tilde{\chi}}^0_1\tilde{\chi}^0_1 G_{\mu\nu}^aG^{\mu\nu}_a
	\label{eq:L_chi_q}
\end{align}
where $\lambda_{q},\lambda_g, g_q^{(i)}$ are the corresponding Wilson coefficients, $G_{\mu\nu}^a$ is the gluon field strength tensor and $m_{\tilde\chi_1^0}$ is the LSP ($\tilde{\chi}_1^0$) mass. The second and third term in $\mathscr{L}_q^{\rm el}$  involves quark \textbf{twist-2} operator $\mathcal{O}^q_{\mu\nu}$~\cite{Drees1, Drees2}.
The effective neutralino-nucleon { effective} coupling, $f_N$ can be evaluated from parton level Lagrangian as~\cite{Hisano:2004pv, Bisal:2024iar,Hisano:2011cs},
\begin{align}
	\frac{f_N^{\rm el}}{m_N}=\sum_{u,d,s}\lambda_qf_q^N+\sum_{u,d,s,c,b}\left(\frac34\left[q(2)+\bar q(2)\right]\left(g_q^{(1)}+g_q^{(2)}\right)\right)-\frac{8\pi}{9\alpha_s}\lambda_gf_g^N
\end{align}
where $f_q^N,f_g^N$ are the quark and gluon form factors for nucleon $N\in\{n,p\}$, with mass $m_N$, and $q(2)$ and $\bar q(2)$ are the second moments of the parton distribution functions (PDFs)~\cite{Hisano:2011cs}. {The above effective coupling as well as the elastic cross-section are evaluated using {\tt micrOMEGAs}, which includes the \textbf{twist-2} contributions at leading order.} 

\subsection{Inelastic neutralino-nucleon spin-independent scattering}
Similar to the scalar operator in the elastic scattering, the inelastic scattering can occur through a different set of scalar operators given as, 
\begin{align}\label{eq:Leff}
	\mathcal{L}_ {\rm eff}^{\rm inel} \supset \sum_{u,d,s}\lambda_q' \bar{\tilde{\chi}}^0_1\tilde{\chi}^0_2
	\bar{q}q +\lambda_g'\bar{\tilde{\chi}}^0_1\tilde{\chi}^0_2 G_{\mu\nu}^aG^{\mu\nu}_a+\dots
\end{align}
The effective { $\tilde\chi_1^0\tilde\chi_2^0\bar NN$} SI coupling can be obtained from the above parton-level Lagrangian as,
\begin{align}\label{eq:higgs_nucleon}{
    \frac{f_N^{\rm inel}}{m_N}=\sum_{u,d,s}\lambda_q'f_q^N-\frac{8\pi}{9\alpha_s}\lambda_g'f_g^N}
\end{align}
{ Note that, for simplicity, we consider scalar contributions to the inelastic effective coupling only through Higgs exchanges, unlike the elastic case, where additional contributions from other diagrams (\textit{e.g.} twist-2, squark mediation \textit{etc.}) are included within {\tt micrOMEGAs}. Additionally, we only consider dominant heavy SM quark ($c,b,t$) loops in the effective $ggh_j$ coupling as the contributions from other SM quarks and MSSM squarks are supressed due to small Yukawa and relatively heavier mass, respectively.}
The minimum DM velocity required to excite to the heavier state $\tilde\chi_2^0$ necessary for inelastic scattering, along with imparting a recoil energy $E_R$ to the target nucleus, is given by Eq.~(\ref{eq:vminInel}) \cite{Tucker-Smith:2001myb} (see \textbf{Appendix }\eqref{Kinematics} for details). 

\section{Radiative Corrections to $\tilde\chi_1^0 \tilde{\chi}_i^0 h_j$ vertices}
\label{sec:Rad Corr}
While the $\tilde\chi_1^0 \tilde{\chi}_i^0 h_j$ vertices 
required for DD are present at tree level, they can be vanishingly small in certain parts of the MSSM parameter space, such as in pure Bino or Higgsino scenarios. 
Thus, we compute the one-loop electroweak corrections {to electroweakino masses and $\tilde{\chi}_1^0 \tilde{\chi}_i^0 h_j$ vertices} using dimensional regularization and the on-shell renormalization scheme. 
For the precise calculations of the Higgs masses and the other SUSY particles we use {\tt SPheno-4.0.4}~\cite{Porod:2003um, Staub:2017jnp}, which uses model files from {\tt SARAH-4.14.5}~\cite{STAUB20141773, Staub:2017jnp, Staub:2015kfa}.

\subsection{EW corrections to $\tilde\chi_1^0\tilde\chi_i^0h_j$ vertices}

In MSSM, in the mass eigenbasis, the tree-level Lagrangian involving
$\tilde\chi_1^0 \tilde\chi_2^0 h_j$ vertices can be written as
\cite{Drees:2004jm},
\begin{equation}
	\mathcal{L}  \supset  \sum_{j\in\{1,2\}} \left[- \dfrac{1}{2} h_j \bar{\tilde{\chi}}_1^0 ({\mathscr{C}}^{L}_{j}
    \mathbf{P_L} + {\mathscr{C}}^{R}_{j} \mathbf{P_R}) \tilde{\chi}_1^0   -
    \dfrac{1}{2}    h_j \bar{\tilde{\chi}}_1^0
	({\mathcal{C}}^{L}_{j} \mathbf{P_L} + {\mathcal{C}}^{R}_{j} \mathbf{P_R}) \tilde{\chi}_2^0\right]~,
\label{eq:hchi_lgrng}
\end{equation}
where $\mathbf{P_L},\mathbf{P_R}$ are the left and right chirality projection operators. Here, $h_1$ denotes the Standard Model-like Higgs boson with mass $\sim125$ GeV, while
$h_2$ refers to the heavier CP-even scalar. 
The ${\tilde{\chi}}_1^0 {\tilde{\chi}}_i^0h_j$ couplings 
hold ${\mathbf{C}}^{L}_{j} = {\mathbf{C}}^{R*}_{j}~(\mathbf{C}_j\in\{\mathcal{C}_j,\mathscr{C}_j\})$, with the coefficients
shown in Table~\ref{tab:Lag}. 
\begin{table}[h!]
    \centering
    \renewcommand{\arraystretch}{1.4}
    \caption{Lagrangian vertex factors for the $\tilde{\chi}_1^0 \tilde{\chi}_i^0h_j$
    vertices.}
    \begin{tabular}{lccc}
       \hline
       Vertex  & Symbol & $\mathbf{C}_1^{L}(=\mathbf{C}_1^{R*})$ & $\mathbf{C}_2^{L}(=\mathbf{C}_2^{R*})$ \\
       \hline
       $\tilde{\chi}_1^0 \tilde{\chi}_1^0h_j$  ~~& ~~$\mathscr{C}_{j}
       $ ~~&~~ $S_1 \sin\alpha + S_2 \cos\alpha$ ~~&~~ $S_2
       \sin\alpha - S_1 \cos\alpha$ ~~\\
       $\tilde{\chi}_1^0 \tilde{\chi}_2^0h_j$~~  & ~~$\mathcal{C}_{j}
       $ ~~&~~ $Q_1 \sin{\alpha} + Q_2 \cos{\alpha}$ ~~&~~ $Q_2
       \sin\alpha - Q_1 \cos{\alpha}$ ~~\\ 
       \hline  
    \end{tabular}
    \label{tab:Lag}
\end{table}

\begin{figure}[htbp]
			\begin{tikzpicture}
				\begin{feynman}[every edge/.style={line width=0.8pt}]
					\vertex (a1) at (0.3,1.7) {\(\tilde{\chi}^0_1\)};
					\vertex (b1) at (1.5, 1);
					\vertex (b2) at (3.5,1);
					\vertex (b3) at (2.5, -0.5);
					\vertex (a3) at (4.7,1.7) {\(\tilde{\chi}^0_i\)};

					\vertex (c) at (2.5,-1.5);

					\diagram*{
					(a1) -- [fermion] (b1) -- [photon, edge label'=$V$] (b3)-- [photon, edge label'=$V'$] (b2) -- [fermion] (a3),
					(b1) -- [fermion, edge label=$F$] (b2),
					(b3) -- [scalar, edge label=$h_j$] (c),
					};
					\node at (2.5,-2.3) {(a)};
				\end{feynman}
			\end{tikzpicture}
			\begin{tikzpicture}
				\begin{feynman}[every edge/.style={line width=0.8pt}]
					\vertex (a1) at (0.3,1.7) {\(\tilde{\chi}^0_1\)};
					\vertex (b1) at (1.5, 1);
					\vertex (b2) at (3.5,1);
					\vertex (b3) at (2.5, -0.5);
					\vertex (a3) at (4.7,1.7) {\(\tilde{\chi}^0_i\)};

					\vertex (c) at (2.5,-1.5);

					\diagram*{
					(a1) -- [fermion] (b1) -- [fermion, edge label'=$F$] (b3)-- [fermion, edge label'=$F'$] (b2) -- [fermion] (a3),
					(b2) -- [photon, edge label'=$V$] (b1),
					(b3) -- [scalar, edge label=$h_j$] (c),
					};
					\node at (2.5,-2.3) {(b)};
				\end{feynman}
			\end{tikzpicture}
			\begin{tikzpicture}
				\begin{feynman}[every edge/.style={line width=0.8pt}]
					\vertex (a1) at (0.3,1.7) {\(\tilde{\chi}^0_1\)};
					\vertex (b1) at (1.5, 1);
					\vertex (b2) at (3.5,1);
					\vertex (b3) at (2.5, -0.5);
					\vertex (a3) at (4.7,1.7) {\(\tilde{\chi}^0_i\)};

					\vertex (c) at (2.5,-1.5);

					\diagram*{
					(a1) -- [fermion] (b1) -- [scalar, edge label'=$S$] (b3)-- [photon, edge label'=$V$] (b2) -- [fermion] (a3),
					(b1) -- [fermion, edge label=$F$] (b2),
					(b3) -- [scalar, edge label=$h_j$] (c),
					};
					\node at (2.5,-2.3) {(c)};
				\end{feynman}
			\end{tikzpicture}

            \begin{tikzpicture}
				\begin{feynman}[every edge/.style={line width=0.8pt}]
					\vertex (a1) at (0.3,1.7) {\(\tilde{\chi}^0_1\)};
					\vertex (b1) at (1.5, 1);
					\vertex (b2) at (3.5,1);
					\vertex (b3) at (2.5, -0.5);
					\vertex (a3) at (4.7,1.7) {\(\tilde{\chi}^0_i\)};

					\vertex (c) at (2.5,-1.5);

					\diagram*{
					(a1) -- [fermion] (b1) -- [photon, edge label'=$V$] (b3)-- [scalar, edge label'=$S$] (b2) -- [fermion] (a3),
					(b1) -- [fermion, edge label=$F$] (b2),
					(b3) -- [scalar, edge label=$h_j$] (c),
					};
					\node at (2.5,-2.3) {(d)};
				\end{feynman}
			\end{tikzpicture}
			\begin{tikzpicture}
				\begin{feynman}[every edge/.style={line width=0.8pt}]
					\vertex (a1) at (0.3,1.7) {\(\tilde{\chi}^0_1\)};
					\vertex (b1) at (1.5, 1);
					\vertex (b2) at (3.5,1);
					\vertex (b3) at (2.5, -0.5);
					\vertex (a3) at (4.7,1.7) {\(\tilde{\chi}^0_i\)};

					\vertex (c) at (2.5,-1.5);

					\diagram*{
					(a1) -- [fermion] (b1) -- [scalar, edge label'=$S$] (b3)-- [scalar, edge label'=$S'$] (b2) -- [fermion] (a3),
					(b1) -- [fermion, edge label=$F$] (b2),
					(b3) -- [scalar, edge label=$h_j$] (c),
					};
					\node at (2.5,-2.3) {(e)};
				\end{feynman}
			\end{tikzpicture}
			\begin{tikzpicture}
				\begin{feynman}[every edge/.style={line width=0.8pt}]
					\vertex (a1) at (0.3,1.7) {\(\tilde{\chi}^0_1\)};
					\vertex (b1) at (1.5, 1);
					\vertex (b2) at (3.5,1);
					\vertex (b3) at (2.5, -0.5);
					\vertex (a3) at (4.7,1.7) {\(\tilde{\chi}^0_i\)};

					\vertex (c) at (2.5,-1.5);

					\diagram*{
					(a1) -- [fermion] (b1) -- [fermion, edge label'=$F$] (b3)-- [fermion, edge label'=$F'$] (b2) -- [fermion] (a3),
					(b2) -- [scalar, edge label'=$S$] (b1),
					(b3) -- [scalar, edge label=$h_j$] (c),
					};
					\node at (2.5,-2.3) {(f)};
				\end{feynman}
			\end{tikzpicture}
\caption{Feynman diagrams contributing to the one-loop radiative corrections. Here, $F$ denotes fermions, $S$ denotes scalar particles, and $V$ denotes vector bosons, and $i,j=1,2$.}
	\label{fig:diagrams}
\end{figure}
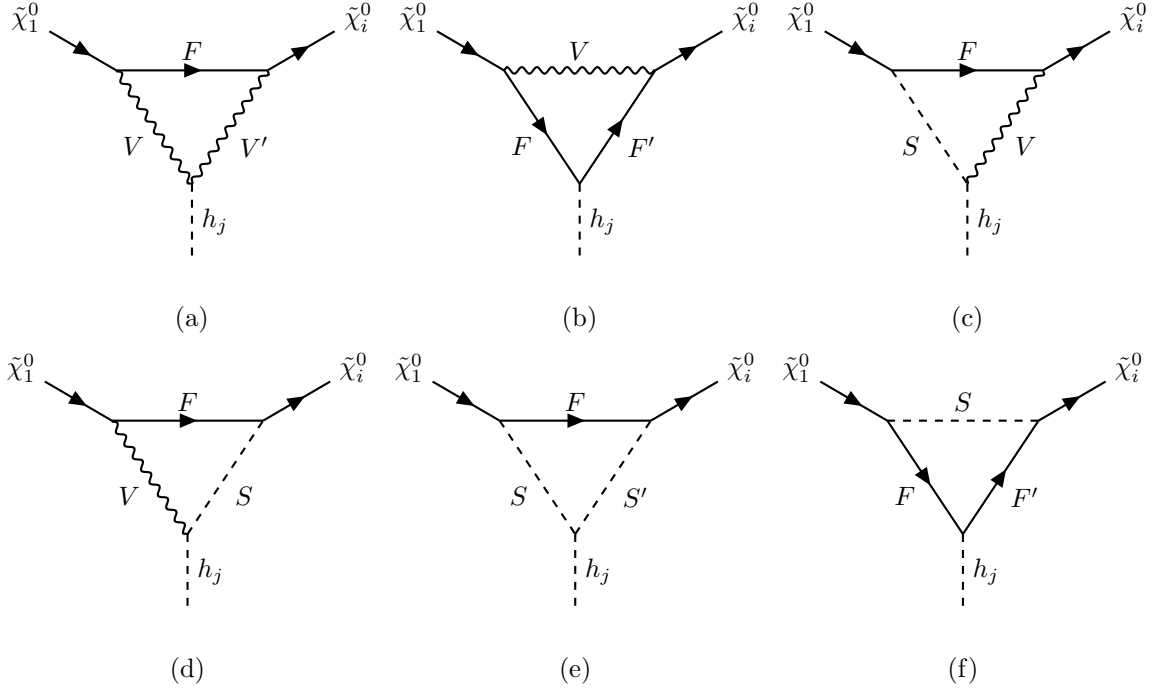

In Table~\ref{tab:Lag}, the symbols have the meaning as follows,
\bea
S_{1} & = &  g_2 N_{13}(N_{12}-\tan\theta_W N_{11}),  \\
S_{2} & = &  g_2 N_{14}(N_{12}-\tan\theta_W N_{11})\,, \\
Q_{1} & = &  g_2 \left[N_{13}(N_{22}-\tan{\theta_W} N_{21})+N_{23}(N_{12}-\tan{\theta_W} N_{11})\right],
\label{eq:Q1}\\
Q_{2} & = &  g_2 \left[N_{14}(N_{22}-\tan{\theta_W} N_{21})+N_{24}(N_{12}-\tan{\theta_W} N_{11})\right]\,.
\label{eq:hchi_coupling}
\eea
The coupling $g_2$ is the $SU(2)$ gauge
coupling, and $\alpha$ parametrizes the mixing in the CP-even Higgs sector. The neutralino
mixing matrix elements $N_{11}$ and $N_{12}$ describe the bino and wino fractions of
$\tilde{\chi}_1^0$, whereas $N_{13}$ and $N_{14}$ correspond to the Higgsino components
associated with the down- and up-type Higgs fields. 
For Higgsino-like $\tilde{\chi}_1^0$ LSP, the gaugino fraction is tiny. As a result,
the tree-level $\tilde{\chi}_1^0 \tilde{\chi}_i^0 h_j$ interactions are strongly suppressed. As shown earlier,
loop-induced corrections play a crucial role in shaping the spin-independent {elastic
scattering} relevant for direct detection experiments \cite{Bisal:2023fgb, Bisal:2024ezn}.
In the same spirit, to accommodate the inelastic contributions to DM-nucleon
 cross-section, we generate
the { $\tilde\chi_1^0 \tilde{\chi}_2^0h_j$ vertices (along with $\tilde\chi_1^0
\tilde{\chi}_1^0h_j$)} at one-loop level through various diagrams, as shown in
Fig.~\ref{fig:diagrams}. The bare Lagrangian can be split into the renormalized and
counterterm (CT) parts as,
\beq
\mathcal{L} = \mathcal{L}_{\rm Born} + \mathcal{L}_{\rm CT}\,.
\eeq
The above parts can be obtained by splitting the ``bare" fields into the renormalized
mass eigenstates and field strength renormalization CTs ($\delta Z$) as,
\begin{align}\label{eq:ren fields}
\tilde{\chi}_i^{0~\rm bare} &= \big(\delta_{ij} + \dfrac{1}{2} \delta {\rm Z}_{ij} \mathbf{P_L}
+ \dfrac{1}{2}
\delta {\rm Z}_{ij}^* \mathbf{P_R} \big) \tilde{\chi}_j^{0~\rm ren},\\
h_i^{\rm bare} &= \big(\delta_{ij} + \dfrac{1}{2} \delta {\rm Z}^{H}_{ij} \big)
h_{j}^{\rm ren},
\label{eq:H field ren}
\end{align}
The wave-function renormalization CTs $\delta Z_{ij}$ are determined using the on-shell renormalization schemes \cite{Fritzsche:2002bi}, a comparison among different variants of the scheme can be found in  Ref.~\cite{Chatterjee:2011wc}.

With the above relations, the counterterm Lagrangian $\mathcal{L}_{\rm CT}$ relevant for
the $\tilde{\chi}_1^0 \tilde{\chi}_i^0 h_j$ vertex can be obtained as follows
\begin{equation}
	-\mathcal{L}_{\rm  CT}   \supset    \sum_{j\in\{1,2\}} \left[- \dfrac{1}{2} h_j \bar{\tilde{\chi}}_1^0 ({\delta\mathscr{C}}^{L}_{j}
    \mathbf{P_L} + {\delta\mathscr{C}}^{R}_{j} \mathbf{P_R}) \tilde{\chi}_1^0   -
    \dfrac{1}{2}    h_j \bar{\tilde{\chi}}_1^0
	({\delta\mathcal{C}}^{L}_{j} \mathbf{P_L} + {\delta\mathcal{C}}^{R}_{j} \mathbf{P_R}) \tilde{\chi}_2^0\right]~,   \nonumber \\
\end{equation}
where $\delta\mathscr{C}_j$ and $\delta\mathcal{C}_j$ are the CTs for $\tilde{\chi}_1^0\tilde{\chi}_1^0 h_j$ and $\tilde{\chi}_1^0\tilde{\chi}_2^0 h_j$ vertices, respectively.
Throughout this work, the CT Lagrangian is written in terms of renormalized fields, and the corresponding renormalization label is omitted for brevity.

\subsection{EW corrections to Electroweakino masses}
As mentioned earlier, the inelastic scattering process involves lightest charginos ($\tilde\chi_1^\pm$) and second lightest neutralino ($\tilde\chi_2^0$), in addition to the LSP ($\tilde\chi_1^0$). In this section, we calculate one-loop corrected masses of these particles in the on-shell scheme. The loop corrected masses can be written as,{
\begin{align}\label{eq:NeuOS}
    m_{\tilde{\chi}^{0}_n} = m_{\tilde{\chi}^{0}_n}^{\rm tree} +
    \left(N^* \delta M^{\rm n} N^{-1}\right)_{nn} - \delta
    m^{\mathrm{OS}}_{\tilde{\chi}^{0}_n}
\end{align}
and
\begin{align}\label{eq:ChaOS}
    m_{\tilde{\chi}^{\pm}_c} = m_{\tilde{\chi}^{\pm}_c}^{\rm tree} +
    \left(U^* \delta M^{\rm c} V^{-1}\right)_{cc} - \delta
    m^{\mathrm{OS}}_{\tilde{\chi}^{\pm}_c}
\end{align}
where $N_{ij}$ are the neutralino mixing matrix elements, $M^{\rm n}$ is the
neutralino mass matrix defined in \cite{Chatterjee:2011wc}, $U$ and $V$
are the chargino mixing matrices and $M^{\rm c}$ is the
chargino mass matrix \cite{Chatterjee:2011wc}.} $m^{\rm tree}$ is the tree
level masses with $\delta m^{\rm OS}$ as corresponding one-loop mass correction and can
be evaluated as,
{
\begin{align} \label{eq:NeuOS1}
\delta m^{\mathrm{OS}}_{\tilde{\chi}^0_n}
&=
\widetilde{\mathrm{Re}}
\left[
m_{\tilde{\chi}^0_n}\,
\Sigma^{L}_{\tilde{\chi}^0}\!\left(m_{\tilde{\chi}^0_n}^{2}\right)
+
\Sigma^{SL}_{\tilde{\chi}^0}\!\left(m_{\tilde{\chi}^0_n}^{2}\right)
\right]_{nn},\\
\delta m^{\mathrm{OS}}_{\tilde{\chi}^\pm_c}
&=
\widetilde{\mathrm{Re}}
\left[
\frac{m_{\tilde{\chi}^\pm_c}}{2}
\left(
\Sigma^{L}_{\tilde{\chi}^\pm}\!\left(m_{\tilde{\chi}^\pm_c}^{2}\right)
+
\Sigma^{R}_{\tilde{\chi}^\pm}\!\left(m_{\tilde{\chi}^\pm_c}^{2}\right)
\right)
+
\Sigma^{SL}_{\tilde{\chi}^\pm}\!\left(m_{\tilde{\chi}^\pm_c}^{2}\right)
\right]_{cc}
\label{eq:ChaOS1}
\end{align}
where 
$\widetilde{\rm Re}$ takes the real part of the loop integrals, but leaves complex couplings
unaffected \cite{Chatterjee:2011wc, Fritzsche:2013fta}. The matrices $\Sigma^L, \Sigma^{SL}$
are the left-handed vector and scalar self-energy corrections, respectively, calculated at $p^2=m^2_{\tilde\chi_i^{0,\pm}}$.
It may be noted that the subscripts ``$nn$" and ``$cc$" in Eqs.~(\ref{eq:NeuOS}--\ref{eq:ChaOS})
denote that only the diagonal two-point functions contribute to the corrections to physical
masses at one-loop level in the on-shell renormalization scheme \cite{Chatterjee:2011wc,
Fritzsche:2002bi}.
{Since the \texttt{CCN[n]} scheme has been used to evaluate the mass corrections
(where $n=3,4$), both the charginos and one of the neutralinos ($\tilde{\chi}_n^0$)
will not receive any corrections as the self-energy corrections will cancel with the
corresponding counterterms \cite{Chatterjee:2011wc} (the second terms of Eqs.
(\ref{eq:NeuOS}-\ref{eq:ChaOS})), i.e.,
$$
\delta m^{\mathrm{OS}}_{\tilde{\chi}^{\pm}_c} = \left(U^*
\delta M^{\rm c} V^{-1}\right)_{cc},
$$ and
$$
\delta m^{\mathrm{OS}}_{\tilde{\chi}^{0}_n} = \left(N^*
\delta M^{\rm n} N^{-1}\right)_{nn}.
$$}
}


\section{Improvements beyond \texttt{micrOMEGAs} with loop corrected parameters using \texttt{FeynArts}}
\label{sec:Improv}
We outline below the procedure adopted to compute the radiative corrections.

\begin{itemize}

\item The particle spectrum is obtained using \texttt{SPheno-4.0.4}~\cite{Porod:2003um, Staub:2017jnp} with MSSM model files from \texttt{SARAH-4.14.5}~\cite{STAUB20141773, Staub:2017jnp, Staub:2015kfa}. The package calculates 2-loop Higgs masses and both tree- and 1-loop masses for all the MSSM particles in $\overline{DR}$ renormalization scheme. 

\item We compute the one-loop corrections to the masses of $\tilde\chi_1^0$, $\tilde\chi_2^0$,
and $\tilde\chi_1^\pm$ {(equivalently, $\delta$ and $\delta m_{\tilde{\chi}_1^{\pm}}$)} in the on-shell scheme using Eqs.(\ref{eq:NeuOS}-\ref{eq:ChaOS1}). The amplitudes are generated using
\texttt{FeynArts-3.11}~\cite{KUBLBECK1990165, HAHN2001418} and reduced to Passarino--Veltman
(PV) functions with \texttt{FormCalc-9.10}~\cite{PASSARINO1979151}. The resulting
expressions, together with the corresponding counterterms, are exported to {\tt FORTRAN}
code. The numerical evaluation for each parameter point is performed using
\texttt{LoopTools-2.16}~\cite{HAHN1999153}. The corrected masses are then used to update
the MSSM spectrum file.

\item In a similar manner, we generate the LO and 1-loop amplitudes for the $\tilde{\chi}_1^0\tilde{\chi}_1^0 h_j$ and $\tilde{\chi}_1^0\tilde{\chi}_2^0 h_j$ processes
using \texttt{FeynArts-3.11}~\cite{KUBLBECK1990165, HAHN2001418} and \texttt{FormCalc-9.10}~\cite{PASSARINO1979151}, and perform the numerical evaluation with \texttt{LoopTools-2.16}~\cite{HAHN1999153}. Moreover, corresponding amplitudes for the counterterm diagrams
are added to the 1-loop amplitudes to get a finite result for the $\tilde{\chi}_1^0
\tilde{\chi}_i^0 h_j$ vertices at NLO.
{The renormalization constants for CTs are evaluated taking both the charginos ($\tilde\chi_{1,2}^\pm$) and one of the 3rd or 4th neutralino ($\tilde\chi_n^0$) to be on-shell. This scheme is referred to as the {\tt CCN[$n$]} scheme in {\tt FeynArts}
\cite{Fritzsche:2002bi, Fritzsche:2011nr, Fritzsche:2013fta}, where `\texttt{CC}' and
`{\tt N[$n$]}', referring to two charginos and neutralino-$n$, respectively, are kept at the poles of their propagators \cite{Eberl:2001iw, Oller:2003ge, Oller:2005xg, Drees:2006um, Heinemeyer:2011gk, Chatterjee:2012hkk, Bharucha:2012re}. This work follows the
convention of \texttt{FeynArts} 
for the on-shell renormalization.}
The NLO vertex consists of the LO vertex, the renormalized vertex, and the corresponding
counterterm vertex, depicted as 
\begin{align}
    \mathscr{C}^{\text{NLO}}_{L,R} &= \mathscr{C}^{\text{LO}}_{L,R} +
    \mathscr{C}^{1\text{L}}_{L,R} + \delta \mathscr{C}_{L,R} \, \text{, and }\\
    \mathcal{C}^{\text{NLO}}_{L,R} &= \mathcal{C}^{\text{LO}}_{L,R} + \mathcal{C}^{1\text{L}}_{L,R} + \delta \mathcal{C}_{L,R} \,,
\end{align}
for the $\tilde{\chi}_1^0\tilde{\chi}_1^0 h_j$ and $\tilde{\chi}_1^0\tilde{\chi}_2^0 h_j$
vertices, respectively.

\item We generate model files for \texttt{micrOMEGAs-6.3.0}~\cite{Belanger:2008sj,Belanger:2004yn,Belanger:2006is, Belanger:2010pz} using {\tt SARAH-4.14.5}~\cite{Staub:2013tta,Staub:2017jnp, Staub:2015kfa}. The $\tilde{\chi}_1^0\tilde{\chi}_1^0 h_j$ vertex is then replaced by its one-loop corrected form, which is subsequently used by \texttt{micrOMEGAs} to evaluate the amplitudes and cross-sections for elastic DM--nucleon scattering.

\item The inelastic DM--nucleon amplitudes are computed using our own routines, based on Eq.\eqref{eq:higgs_nucleon}.


\item The differential LSP--nucleus cross-section is subsequently computed using Eq.\eqref{2}. Subsequently, the differential event rate is obtained from Eq.\eqref{1}. Finally, the total number of events are evaluated using Eq.\eqref{eq:nEvents} and compared against the null result of LZ experiment ($N_E\le\log10$ at 90\% \emph{C.L.}). 

\end{itemize}


\section{Results and Discussion}
\label{sec:results}
In this section, we first outline the objectives that guide our analysis, followed by the presentation of results and their implications.

\begin{itemize}
\item We identify the region of parameter space corresponding to a Higgsino-like neutralino LSP,
 with an one-loop mass splitting $\delta\lesssim\mathcal{O}(100\,\text{keV)}$,
such that the inelastic scattering process becomes comparable in significance to the elastic scattering channel.
{For such small mass splittings, the recoil energy signal for an inelastic collision
will be similar to the elastic one up to a factor proportional to the square of
$\tilde\chi_1^0\tilde\chi_2^0h_j$ couplings.}

\item Consequently, we compute the direct detection cross-sections and corresponding event rates for both elastic and inelastic scattering processes involving the NLO corrected $\tilde\chi_1^0\tilde\chi_i^0h_j$
vertices within this parameter region.

\item We perform a consistent comparison with DD experimental results by:
\begin{itemize}
    \item evaluating elastic and inelastic scattering event rates independently (see Eq.~(\ref{eq:rate}) in the appendix),
    \item combining these contributions to obtain the total event rate (see Eq.~(\ref{eq:nEvents})), and
    \item comparing the total predicted { events with null results at $90\%$ C.L. }($N_{E}^{\rm theory}< \log10\approx2.3$).
\end{itemize}

\item We check suppression effects specific to inelastic scattering, involving:
\begin{itemize}
    \item phase space suppression induced by the mass splitting $\delta$, as reflected in $v_{\rm min}^{\rm inel}$ (Eq.~(\ref{eq:vminInel})),
    \item kinematic constraints arising from recoil energy thresholds, notably the condition $(E_R)_{\rm min} > 0$.
\end{itemize}

\item Finally, we investigate whether the inelastic rate can exceed the elastic rate to be the dominant DD channel in some parameter space region. 

\end{itemize}

We begin by identifying regions of the parameter space that exhibit a small mass splitting $\delta$ between the neutral Higgsino states, such that inelastic scattering is not Boltzmann suppressed and can contribute significantly to the differential event rate $dR/dE_R$ (see Eq.\eqref{1}) in the detector. Additionally, we impose that the $\tilde{\chi}_1^\pm$--$\tilde{\chi}_1^0$ mass splitting, $\delta m_{\tilde{\chi}_1^{\pm}}$, is larger than the electron mass ($m_e = 0.5~\text{MeV}$). This ensures that the chargino can decay promptly, thereby avoiding constraints from stable charged particles at the LHC~\cite{ATLAS:2022pib}. As a consequence, the inelastic process $\tilde{\chi}_1^0 N \to \tilde{\chi}_1^\pm N$ is Boltzmann suppressed and can be safely neglected. We perform a scan over the parameter space for four representative values of the Higgsino mass parameter $\mu$, while varying the gaugino mass parameters $M_1$ and $M_2$ as follows,
\begin{align}
    \mu=\pm 0.5& \,\rm TeV,~\pm1\,\rm TeV\nonumber\\
   {-21 \,\rm TeV}\phantom{-}\leq &\quad M_1 \quad\leq -1 {\rm~\TeV,} \nonumber\\
    1 \,\rm TeV\phantom{-}\leq &\quad M_2 \quad\leq \phantom{-}21 {\rm~\TeV.}\label{eq:scan_range}
\end{align}
with the rest of the parameters fixed at values listed in Table~\ref{tab:inputs_transposed}.

\begin{table}[h!]
\renewcommand{\arraystretch}{1.3}
\centering
\caption{Fixed input parameters for the scan in the range provided in Eq.\eqref{eq:scan_range}.}
\begin{tabular}{ccccccccc}
\hline
$\tan\beta$ &$T_t$ & $m_{\tilde{Q}_L}$ & $m_{\tilde{t}_R}$ & $m_{\tilde{b}_R}$ &
$m_{\tilde{L}_L}$ & $m_{\tilde{e}_R}$ &  $m_A$ & $M_3$ \\[2pt]
\hline
$10$ & $-4~\text{TeV}$ & {$11~\text{TeV}$} & {$4~\text{TeV}$} &
$12~\text{TeV}$ & {$10~\text{TeV}$} & {$10~\text{TeV}$} &
{$6~\text{TeV}$} & {$5~\text{TeV}$} \\
\hline
\end{tabular}
\label{tab:inputs_transposed}
\end{table}

\begin{figure}[h!]
\centering
        \includegraphics[width=1\linewidth]{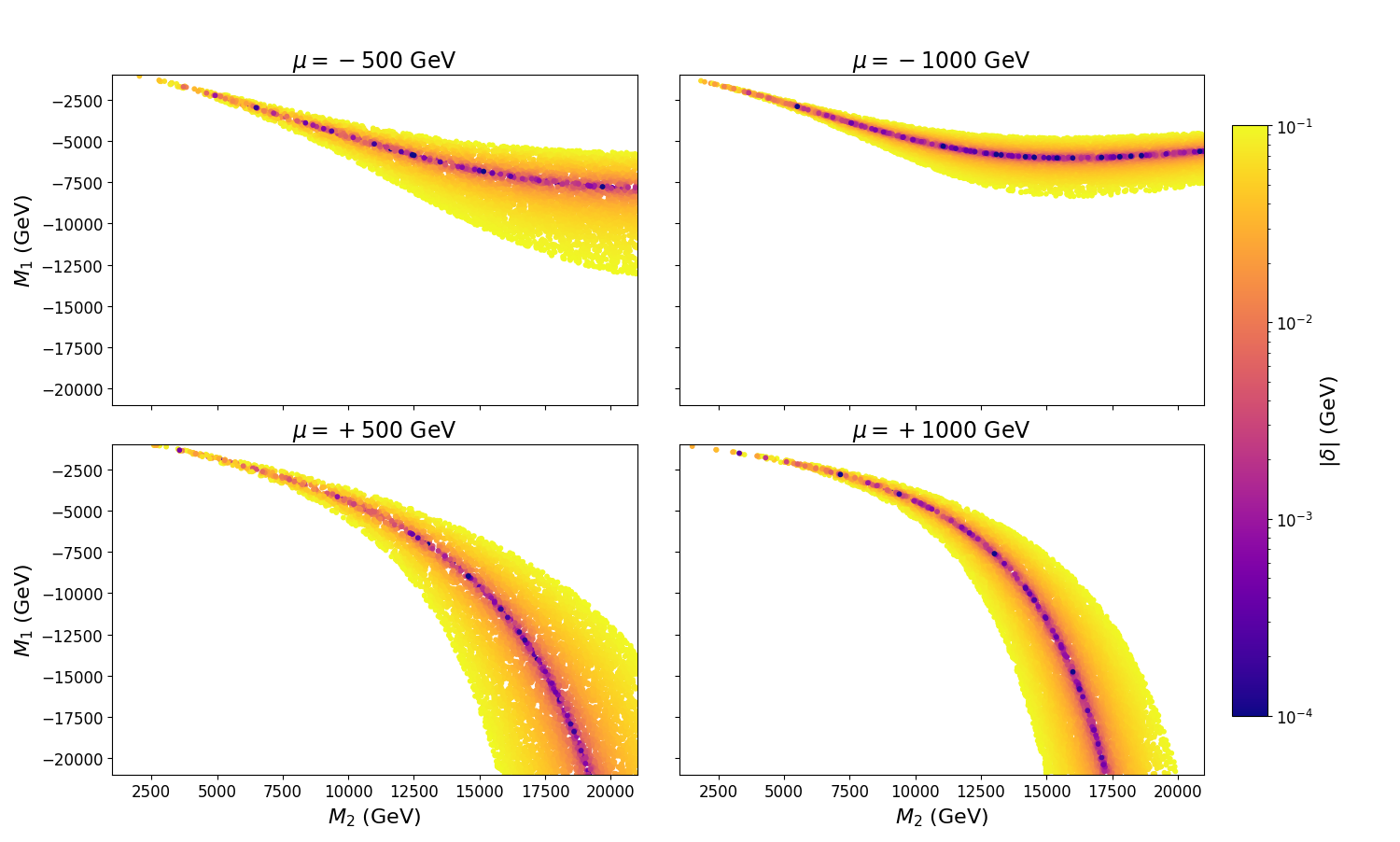}
		\caption{Scattered points with $\delta<0.1$\,GeV, out of the scans in $M_1$--$M_2$ plane (Eq.\eqref{eq:scan_range}), with rest of the parameters fixed at values in Table \ref{tab:inputs_transposed}. The panels correspond to $\mu$ value fixed at (top left) $-500$\,GeV, (bottom left) $500$\,GeV, (top right) $-1000$\,GeV, and (bottom right) 1000\,GeV. Each point is color coded according to the corresponding $\delta$ value as shown in the colorbar.
  		}
\label{fig:neutralino}
\end{figure}

\begin{figure}[h!]
	\centering
	\includegraphics[width=1\linewidth]{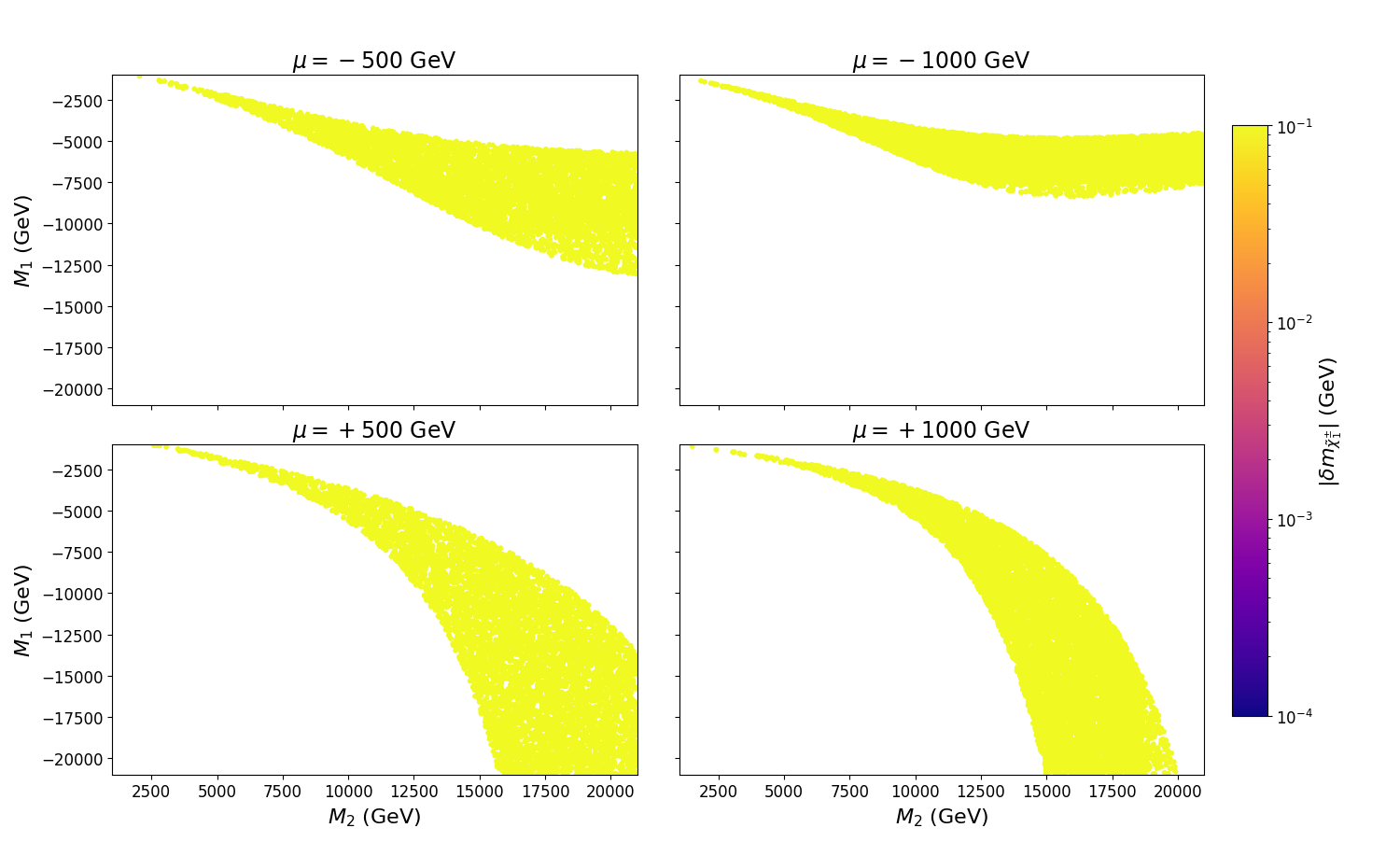}
    \caption{Same points as in Fig.\,\ref{fig:neutralino} but with colors signifying chargino mass splitting, $\delta m_{\tilde{\chi}_1^{\pm}} = m_{\tilde{\chi}_1^{\pm}}-m_{\tilde{\chi}_1^0}$.}
	\label{fig:chargino}
\end{figure}

Fig.~\ref{fig:neutralino} shows the scattered points in the $M_1$–$M_2$ plane satisfying $\delta \leq 0.1$\,GeV, for each representative value of $\mu$, as indicated in the respective panels. The color of each point corresponds to the value of $\delta$, as shown in the color bar. The blue region corresponds to a tiny splitting ($\delta \lesssim \mathcal{O}(100~\text{keV})$), where inelastic scattering can contribute non-negligibly. 
The lightest chargino-LSP 
mass splitting $\delta m_{\tilde{\chi}_1^{\pm}}$
for the same parameter region in the $M_1$--$M_2$ plane as in Fig.~\ref{fig:neutralino}
has been illustrated in Fig.~\ref{fig:chargino}
where, as mentioned, we impose 
$\delta m_{\tilde{\chi}_1^{\pm}} > m_e$ which naturally ensures $\delta m_{\tilde{\chi}_1^{\pm}} > \delta$ for small $\delta$.

\begin{figure}[h!]
	\centering

    \includegraphics[width=0.55\linewidth]{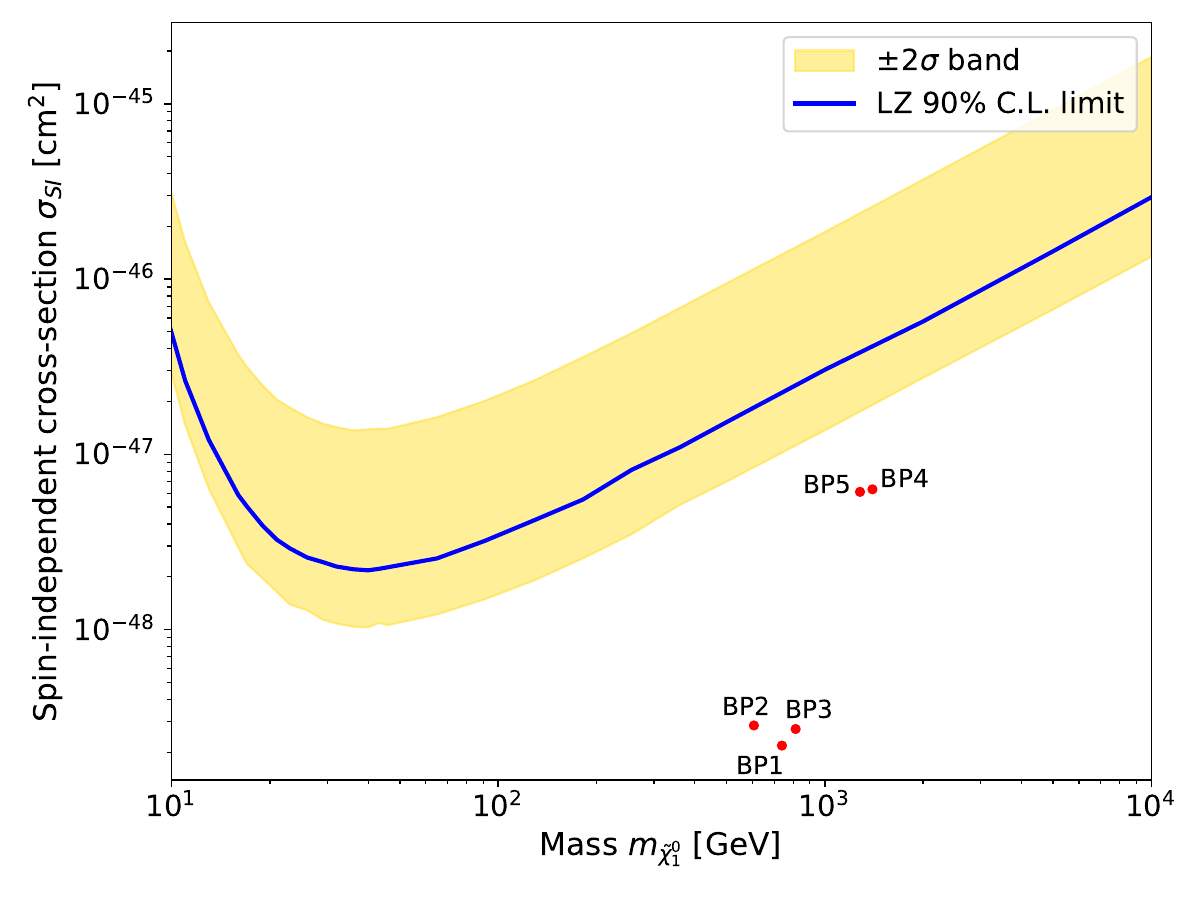}
	\caption{Benchmark points comparison with the LZ experiment \cite{LZCollaboration:2024lux}
		for LO elastic scattering cross-sections.
        }
	\label{fig:exclusion}
\end{figure}

\begin{figure}[h!]
	\centering
	\includegraphics[width=0.48\linewidth]{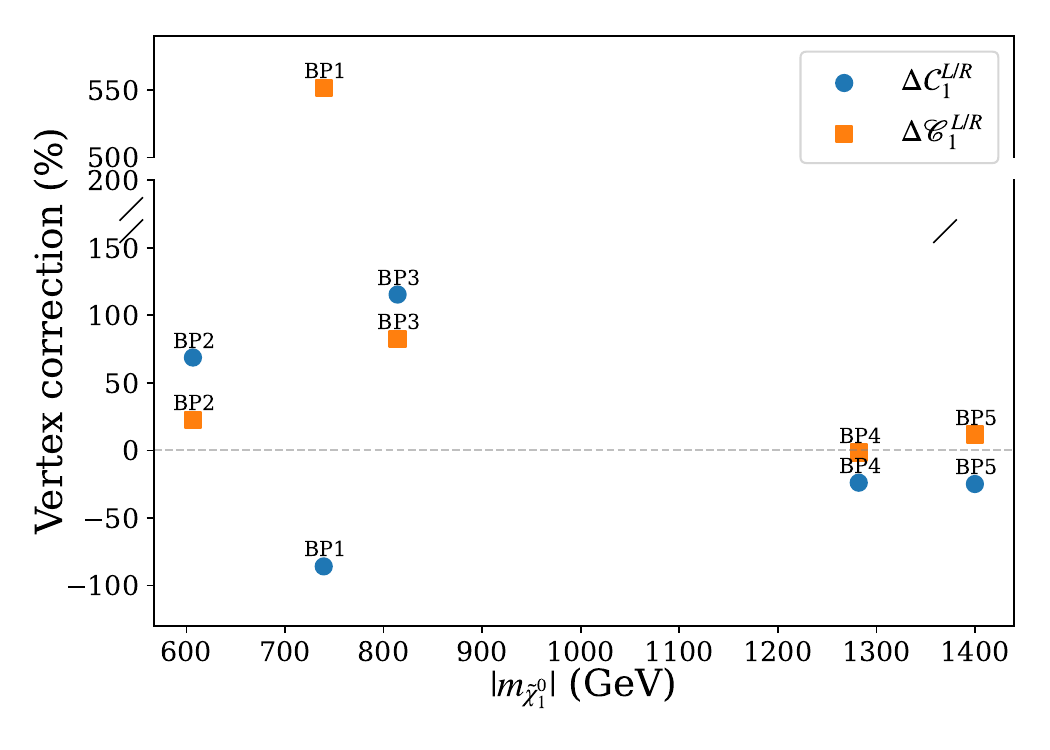}
	\includegraphics[width=0.48\linewidth]{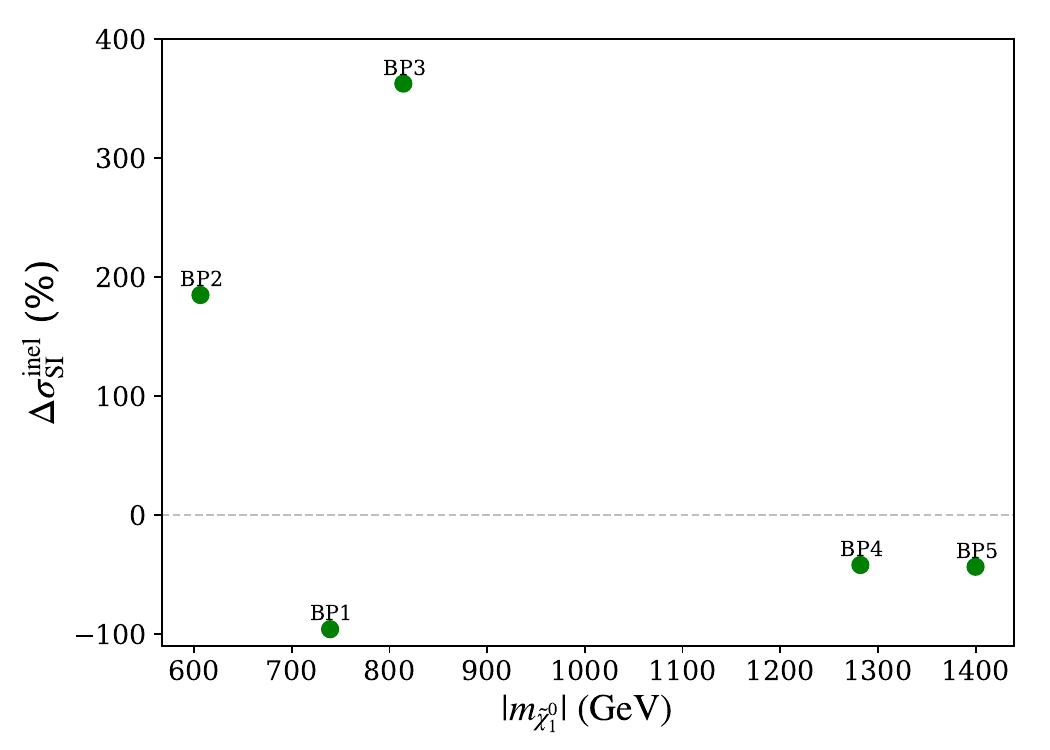}\\
	(a) \hspace{7cm} (b) \\[1pt]
	\caption{ Panel (a) depicts percentage corrections as listed in Table \ref{tab:Results}, to $\tilde{\chi}_1^0 \tilde{\chi}_2^0 h_1$ ($\mathcal{C}_{1}^{\rm L/R}$) and $\tilde{\chi}_1^0 \tilde{\chi}_1^0 h_1$  ($\mathscr{C}_{1}^{\rm L/R}$) vertices at NLO. Panel (b) shows the percentage NLO corrections to the inelastic cross-section ($\sigma_{\rm SI}^{\rm inel}$).}
	\label{fig:corrections}
\end{figure}

Identifying the parameter regions with low $\delta$ guides us in selecting five benchmark points (BPs) from different regions of the parameter space for further analysis. 
For a valid BP, (I)
We restrict the mass of the light Higgs state between $122\,{\rm GeV}\le m_{h_1}\le128$\,GeV, (II)
test the Higgs spectra using {\tt HiggsTools-1.1.3}~\cite{Bahl:2022igd,Bechtle:2020pkv,Bechtle:2020uwn}, and, (III)
check the viability against LHC constraints using {\tt SModelS-3.1.0}
\cite{Altakach:2024jwk, MahdiAltakach:2023bdn, Alguero:2021dig,
Kraml:2013mwa}
with one-loop corrected masses\footnote{As already mentioned, we do not
consider the thermal relic constraints as a primary criterion to select BPs.}. 
The BPs are listed in Table \ref{tab:bp1}, which includes the input parameters and the
output.
{The benchmark points are so chosen that mass splitting of the on-shell
loop corrected neutralino masses, $\delta$, is of order $\mathcal{O}(10)$ keV,
so that the Boltzmann suppression in the inelastic scattering is reduced.
The physical mass splitting is controlled by the comparatively small difference
between the radiative mass corrections, together with the gaugino-Higgsino mixing
contributions. In particular, as can be seen, for example, from Eq. (9) of
Ref.~\cite{Drees:1996pk}, the corrections proportional to $M_Z^2/M_1$ and $M_Z^2/M_2$
depend on the relative signs of $M_1$ and $M_2$. In the parameter regions relevant
for our analysis, these contributions can have opposite signs and partially cancel.
The difference between the radiative corrections to the two Higgsino-like states can
then further increase or decrease this already small splitting.

Moreover, the $\mu$ parameter is varied to observe the effect of neutralino masses
and the Higgsino-gaugino mixing on the radiative corrections. Furthermore, the
relative signs of $\mu$, $M_1,$ and $M_2$ are varied, $\tan{\beta}$ is varied,
the right-type stop mass $m_{\tilde{t}_R}$ is varied, and the trilinear coupling
for the top squark sector ($T_t$) is varied; all to observe their correlation with the
radiative corrections to the $\tilde\chi_1^0\tilde\chi_i^0h_j$ vertices.}
{Fig.~\ref{fig:exclusion} shows the LO cross-sections for these benchmark points,
compared with the central value (blue line) of the 90\% C.L. exclusion line and the
$2\sigma$ band (yellow) reported by LZ \cite{LZCollaboration:2024lux}.
All benchmark points are consistent with the bounds from the LZ experiment at LO.}
{Fig.~\ref{fig:corrections} displays the percentage corrections to the vertex factors and the inelastic scattering cross-section for each BP. This percentage correction is defined as  
\begin{align} \label{eq:Delta}
\Delta Q&=
\frac{Q_{\rm NLO}-Q_{\rm LO}}{Q_{\rm LO}}
\times 100\%~,\quad\quad
Q\in\{\mathscr C_j^{\rm L/R},~\mathcal C_j^{\rm L/R},~\sigma_{\rm SI}^{\rm inel}\},
\end{align}
where $\mathscr C_j^{\rm L/R}$, and $~\mathcal C_j^{\rm L/R}$ denote the vertex factors to the $\tilde{\chi}_1^0 \tilde{\chi}_1^0 h_j$ and the $\tilde{\chi}_1^0 \tilde{\chi}_2^0 h_j$ vertices, respectively.}

\begin{table}[h!]
\renewcommand{\arraystretch}{1.05}
\centering
\caption{
Parameters and spectrum in the Benchmark scenarios. Here, OS (on-shell) and $\overline{DR}$ signify the renormalized schemes in which the one-loop masses are evaluated. All the sfermions except $m_{\tilde t_R}$ are heavy with $m_{\tilde f}=10$\,TeV. The gluino and pseudoscalar masses are fixed at 5\,TeV and 3\,TeV, respectively. The elastic cross-section at LO is also listed, which is calculated using {\tt MicrOMEGAs-6.3.0}~\cite{Belanger:2008sj}. All the parameters are in GeV unless stated otherwise. { For all BPs $\Omega_{\rm DM}h^2 < 0.12$.} }
\begin{tabular}{clccccc}
\hline
 & Parameter & BP1 & BP2 & BP3 & BP4 & BP5 \\
\hline

\multirow{6}{*}{{Inputs}~}
& $\mu$ & $-739.63$ & $-606.64$ & $-814.26$ & 1282.11 & 1400.0 \\
& $M_1$ & $-7830$ & $-6292$ & $-5838$ & 11354 & 12875 \\
& $M_2$ & 17055 & 42000 & 30925 & 12627 & 12050 \\
& $\tan\beta$ & 10.88 & 15.83 & 15.83 & 25.89 & 26.46 \\
& $m_{\tilde{t}_R}$ & 2290 & 663.32 & 860 & 1510 & 1450 \\
& $T_t$ & $-3660$ & $-4093$ & $-4093$ & $-3720$ & $-3717$ \\
\hline

\multirow{20}{*}{{Outputs}~}
& $m_{\tilde{\chi}_1^0}$ (tree) & 739.55 & $-606.54$ & $-814.17$ & 1281.70 & 1399.59 \\

& $m_{\tilde{\chi}_1^0}$ (1-loop, OS) & 739.33 & $-606.65$ & $-814.25$ & 1282.01 & 1399.84 \\

& $m_{\tilde{\chi}_2^0}$ (tree) & $-739.69$ & 606.69 & 814.29 & $-1282.39$ & $-1400.28$ \\

& $m_{\tilde{\chi}_2^0}$ (1-loop, OS) & $-739.33$ & 606.65 & 814.25 & $-1282.01$ & $-1399.84$ \\

& $m_{\tilde{\chi}_1^\pm}$ & 739.68 & 606.66 & 814.28 & 1282.02 & 1399.90 \\


& $\bm\delta$ (1-loop, OS) [keV] &\bf  36.37 &\bf {8.90} &\bf 16.39 & {\bf 10.81} &\bf 10.59 \\
\cline{2-7}
& $N_{11}\,(\times10^{-3})$ & $-3.20$ & 5.69 & 6.439 & $-3.14$ & $2.755$ \\

& $N_{12}\,(\times10^{-3})$ & $-3.149$ & 1.415 & 1.899 & 5.197 & $-5.531$ \\

& $N_{13}\,(\times10^{-1})$ & $-7.071$ & $-7.071$ & $-7.071$ & $-7.072$ & $7.072$ \\

& $N_{14}\,(\times10^{-1})$ & $-7.071$ & 7.070 & 7.071 & 7.070 & $-7.070$ \\

& HF (\%) & 99.998 & 99.997 & 99.995 & 99.996 & 99.996 \\
\cline{2-7}
& $m_{h_1}$ & 123.12 & 125.78 & 124.40 & 123.67 & 123.67 \\

& $m_{h_2}$ & 7825.2 & 6871.8 & 7070.4 & 8459.8 & 8469.2 \\

& $m_{\tilde{\chi}_3^0}$  (tree) & $-7830.24$ & $-6292.30$ & $-5838.32$ & 11354.16 & 12050.5 \\
& $m_{\tilde{\chi}_4^0}$ (tree)  & 17055.4 & 42000.2 & 30925.2 & 12627.5 & 12875.1 \\
& $m_{\tilde{\chi}_2^\pm}$  & 17055.4 & 42000.2 & 30925.2 & 12627.5 & 12050.5 \\

&$m_{\tilde t_1}$ (tree) & 2294.34 & 680.53 & 873.35 & 1309.08 & 1458.15 \\

&$m_{\tilde t_1}$ (1-loop, $\overline{DR}$) & 2648.58 & 1447.58 & 1572.34 & 1662.04 & 1712.90 \\
\cline{2-7}

&$\sigma_{SI~\rm LO}^{\rm el}$ [$10^{-12}$ pb]
        & 0.218 & 0.284 & 0.271 & 6.109 & 6.317 \\



\hline
\end{tabular}
\label{tab:bp1}
\end{table}

For the BPs, the values of relevant vertices, including NLO corrections, for elastic and inelastic
scattering processes, along with corresponding {percentage corrections for these
vertices, are
presented in Table~\ref{tab:Results} following the convention shown in
Eq.~(\ref{eq:Delta})}. 
As seen from Table~\ref{tab:Results}, the NLO corrections to $\tilde{\chi}_1^0
\tilde{\chi}_i^0 h_j$ vertices may receive a large correction $\geq |\mathcal O(100)|\%$,
thanks to the smallness of the LO value. 
As the radiative corrections to the $\tilde\chi_1^0\tilde\chi_2^0h_1$ vertex is of importance, Table \ref{tab:contri} enlists the dominant 1-loop contributions to the vertex from different diagrams from Fig.\,\ref{fig:diagrams}. Clearly, the total contributions from  $\tilde qqq$-type diagrams are the most dominant for all the BPs, as they include the $\tilde t_1tt$ diagram, which is enhanced due to a larger Yukawa as well as a small $\tilde t_1$ ($\approx\tilde t_R$) mass. This is followed by $\tilde q\tilde qq$ and $\tilde\chi_i^\pm\tilde\chi_j^\pm W^\pm$ type diagrams as shown in Table \ref{tab:contri}.
{This is the reason for the strong dependence of the radiative corrections on the
values of $m_{\tilde{t}_R}$ and $T_t$.}
The contributions from the CP-even heavy Higgs, $h_2$, to the elastic and
inelastic scattering is suppressed compared to $h_1$ as the mass of $h_2$,
$m_{h_2} > 6$ TeV, as compared to $m_{h_1} \sim 125$ GeV. Hence, individual
loop contributions for heavy Higgs are not shown in the table.

\begin{table}[h!]
    \centering
     \caption{
Leading-order (LO) and next-to-leading-order (NLO) couplings for the $\tilde{\chi}_1^0 \tilde{\chi}_1^0 h_j$ ($\mathscr{C}_j^{L/R}$) and $\tilde{\chi}_1^0 \tilde{\chi}_2^0 h_j$ ($\mathcal{C}_j^{L/R}$) vertices, with $j \in \{1,2\}$, along with the corresponding percentage corrections at NLO. 
These are relevant for the evaluation of elastic and inelastic scattering amplitudes. 
}

        \renewcommand{\arraystretch}{1.4}
    \begin{tabular}{lccccc}
\hline
Parameter & BP1 & BP2 & BP3 & BP4 & BP5 \\
        
\hline

        $\mathcal{C}_{1~\rm LO}^{L/R}$ ($\times10^{-4}$)
        & $-5.472$ & 5.713 & 4.581 & 27.115 & 27.39 \\

        $\mathcal{C}_{2~\rm LO}^{L/R}$ ($\times10^{-4}$)
        & $-0.542$ & $-1.959$ & $-2.605$ & $-4.927$ & $-5.214$ \\

        $\mathcal{C}_{1~\rm NLO}^{L/R}$ ($\times10^{-4}$) & -0.772 & 9.640 & 9.866
        & 20.628 & 20.56 \\

        $\mathcal{C}_{2~\rm NLO}^{L/R}$ ($\times10^{-4}$) & $-0.658$ & $-1.738$ & $-2.384$
        & $-2.939$ & $-3.054$ \\

       $\Delta\mathcal{C}_{1}^{L/R}$ (\%) 
        & $-85.90$ & 68.73 & 115.36 & $-23.93$ & $-24.92$ \\

        $\Delta\mathcal{C}_{2}^{L/R}$ (\%) 
        & 21.41 & $-11.29$ & $-8.51$ & $-40.36$ & $-41.42$ \\\hline

        $\mathscr{C}_{1~\rm LO}^{L/R}$ ($\times10^{-3}$)
        & 0.582 & $-0.783$ & $-0.748$ & 3.234 & 3.291 \\

        $\mathscr{C}_{2~\rm LO}^{L/R}$ ($\times10^{-3}$)
        & $-0.700$ & $-0.694$ & $-0.659$ & 2.994 & 3.053 \\

        $\mathscr{C}_{1~\rm NLO}^{L/R}$ ($\times10^{-3}$) & 3.793 & $-0.964$ & $-1.365$
        & 3.184 & 3.678 \\

        $\mathscr{C}_{2~\rm NLO}^{L/R}$ ($\times10^{-3}$) & $-0.312$ & $-1.313$ & $-1.354$
        & 2.608 & 2.595 \\

        $\Delta\mathscr{C}_{1}^{L/R}$ (\%) 
        & 551.31 & $22.42$ & 82.48 & $-1.55$ & 11.75 \\

        $\Delta\mathscr{C}_{2}^{L/R}$ (\%) 
        & $-55.46$ & 89.21 & 105.42 & $-12.91$ & $-14.99$ \\\hline



\hline
    \end{tabular}
    \label{tab:Results}
\end{table}

\begin{table}[h!]
    \centering
    \renewcommand{\arraystretch}{1.4}
    \caption{Finite parts of the dominant loop contributions to the 
$\tilde{\chi}_1^0 \tilde{\chi}_2^0 h_1$ vertex. 
Contributions from the remaining diagrams are subdominant and are therefore not included in this table.}
    \begin{tabular}{lcccccc}
\hline
Loop& Diagram & BP1 & BP2 & BP3 & BP4 & BP5 \\
\hline
$\tilde\chi_i^\pm\tilde\chi_j^\pm W^\pm$ 
& Fig.\,\ref{fig:diagrams}(a)
& $\phantom{-} 2.9 \times 10^{-4}$ 
& $\phantom{-} 1.3 \times 10^{-4}$ 
& $\phantom{-} 1.7 \times 10^{-4}$ 
& $-3.9 \times 10^{-4}$ 
& $-4.1 \times 10^{-4}$ \\

$\tilde{q}_iqq$ 
& Fig.\,\ref{fig:diagrams}(f) 
& $\phantom{-} 4.9 \times 10^{-4}$ 
& $\phantom{-} 5.4 \times 10^{-4}$ 
& $\phantom{-} 6.0 \times 10^{-4}$ 
& $-1.3 \times 10^{-3}$ 
& $-1.3 \times 10^{-3}$ \\

$\tilde{q}_i\tilde{q}_jq$ 
& Fig.\,\ref{fig:diagrams}(e) 
& $-2.5 \times 10^{-4}$ 
& $-5.2 \times 10^{-4}$ 
& $-4.8 \times 10^{-4}$ 
& $\phantom{-} 4.0 \times 10^{-4}$ 
& $\phantom{-} 3.8 \times 10^{-4}$ \\

$\tilde\chi_i^0\tilde\chi_j^0Z$ 
& Fig.\,\ref{fig:diagrams}(a)
& $\phantom{-} 1.35 \times 10^{-5}$ 
& $<10^{-5}$ 
& $<10^{-5}$ 
& $<10^{-5}$ 
& $-2.2 \times 10^{-5}$ \\
\hline
\end{tabular}
    \label{tab:contri}
\end{table}


Summing all the corrections, we compute the LO and NLO cross-sections for both elastic
and inelastic scattering {(at zero momentum transfer limit)}.
Table~\ref{tab:cross-compare} lists the corresponding
numerical values.
{$\sigma_{\rm SI}^{\rm inel}$ ($\sigma_{\rm SI}^{\rm el}$) is the SI proton
cross-section for the inelastic (elastic) processes for $\tilde{\chi}_1^0$-nucleon
scatterings, respectively.}\footnote{For
all BPs, the chargino decay width, $\Gamma(\tilde{\chi}_1^{\pm})$ $\sim 10^{-16}$ GeV.}
For comparison, we also include the ratios of the NLO inelastic
cross-section to the NLO elastic, LO elastic, and LO inelastic cross-sections.
We observe that, although all benchmark points feature a small $\delta$, the inelastic
cross-section for BP1 is subdominant compared to the elastic one. In contrast, for
BP2--BP5, the inelastic cross-section is comparable to the elastic contribution,
highlighting the importance of including these effects in the scattering
event rate in detector materials. For instance, BP2 exhibits an inelastic
cross-section that is approximately a factor of {two larger than the elastic LO
one}.

{The LO cross-sections depend on the tree-level
$\tilde{\chi}_1^0\tilde{\chi}_i^0 h_j$ vertex factors. These couplings are
primarily determined by the Higgsino--Higgs interaction and the Higgsino--gaugino
mixing in $\tilde{\chi}_i^0$. Consequently,
$\sigma_{\rm LO}^{\rm el}$ ($\sigma_{\rm LO}^{\rm inel}$) follows the hierarchy
of $\mathscr{C}^{L/R}_{1\,\rm LO}$ ($\mathcal{C}^{L/R}_{1\,\rm LO}$), with BP1
{(BP3)} yielding the smallest values and BP5 the largest.
At NLO, the elastic (inelastic) cross-section is governed by the corrected
vertex factor $\mathscr{C}^{L/R}_{1\,\rm NLO}$
($\mathcal{C}^{L/R}_{1\,\rm NLO}$). These effective couplings receive
contributions from individual loop diagrams involving quark--squark and
chargino--$W$ boson loops, together with the corresponding counterterm
contributions. The relative sizes and signs of these contributions,
illustrated in Table~\ref{tab:contri} for the
$\tilde{\chi}_1^0\tilde{\chi}_2^0 h_1$ vertex, determine the degree of
constructive or destructive interference, and hence the final NLO vertex
factors. This interplay ultimately gives rise to the NLO cross-sections
reported for the benchmark scenarios in
Table~\ref{tab:cross-compare}.
As shown in the table, the LO elastic contribution in BP1 exceeds the corresponding inelastic contribution. Furthermore, $\sigma_{\rm NLO}^{\rm el}$ receives a sizable, constructively interfering correction, whereas $\sigma_{\rm NLO}^{\rm inel}$ is subject to substantial destructive interference. Consequently, the inelastic scattering rate is strongly suppressed relative to the elastic one in the BP1 benchmark scenario.
On the other hand, in BP2, {the hierarchy at LO is the same, but both have comparable
magnitudes, i.e., $\sigma_{\rm LO}^{\rm el} \approx \sigma_{\rm LO}^{\rm inel}$. More importantly, the radiative corrections enhance the inelastic cross-section more than the corresponding elastic cross-section through constructive interference. As a result, inelastic scattering slightly dominates over elastic scattering in this benchmark.
For BP3, $\sigma_{\rm LO}^{\rm el} > \sigma_{\rm LO}^{\rm inel}$, and both channels receive significant constructive radiative corrections, with the inelastic channel receiving slightly more than the elastic one. Consequently, elastic and inelastic scatterings contribute at a similar level to the total event rate, but the elastic channel dominates.
Finally, for BP4 and BP5, the elastic and inelastic LO cross sections are again of comparable size. However, the inelastic vertex correction, $\Delta\mathcal{C}_{1}^{L/R}$, is negative
for both BP4 and BP5, whereas the elastic vertex correction, $\Delta\mathscr{C}_{1}^{L/R}$, is almost zero for BP4 ($-1.55\%$) and positive for BP5 (11.75\%). This enhances the elastic channel while suppressing the inelastic one, leading to elastic scattering being the dominant contribution. Nevertheless, the inelastic channel remains non-negligible.
}}

\begin{table}[h]
	\centering
	\caption{
    Leading order (LO) and Next-to-leading order (NLO) cross-sections ($10^{-12}$\,pb)
    for elastic ($\sigma_{\rm LO/NLO}^{\rm el}$) and inelastic
    ($\sigma_{\rm LO/NLO}^{\rm inel}$) processes for the Benchmark scenarios of Table
    \ref{tab:bp1} {(in the zero momentum transfer limit)}. The ratio of
    $\sigma_{\rm NLO}^{\rm inel}$ with other cross-sections is also listed for
    comparison.}
    \renewcommand{\arraystretch}{1.4}
    \begin{tabular}{lccccc}
    \hline
     & BP1 & BP2 & BP3 & BP4 & BP5 \\
    \hline
    $\sigma^{\rm el}_{\rm LO}$ 
    & 0.218 & 0.284 & 0.271 & 6.109 & 6.317 \\
    
    $\sigma^{\rm el}_{\rm NLO}$ 
    & 8.122 & 0.504 & 1.021 & 5.595 & 7.503 \\
    
    $\sigma^{\rm inel}_{\rm LO}$ 
    & 0.0869 & 0.178 & 0.120 & 4.754 & 4.878 \\
    
    $\sigma^{\rm inel}_{\rm NLO}$ 
    & 0.0034 & 0.507 & 0.555 & 2.751 & 2.750 \\
    
    ${\sigma^{\rm inel}_{\rm NLO}}/{\sigma^{\rm inel}_{\rm LO}}$ 
    & 0.0391 & 2.848 & 4.625 & 0.579 & 0.564 \\
    
    ${\sigma^{\rm inel}_{\rm NLO}}/{\sigma^{\rm el}_{\rm NLO}}$ 
    & 0.0004 & 1.006 & 0.544 & 0.492 & 0.367 \\
    
    ${\sigma^{\rm inel}_{\rm NLO}}/{\sigma^{\rm el}_{\rm LO}}$ 
    & 0.0156 & 1.785 & 2.048 & 0.4503 & 0.435 \\
    \hline
    \end{tabular}
	\label{tab:cross-compare}
\end{table}

\begin{figure}[h!]
    \centering
    \includegraphics[width=0.7\linewidth]{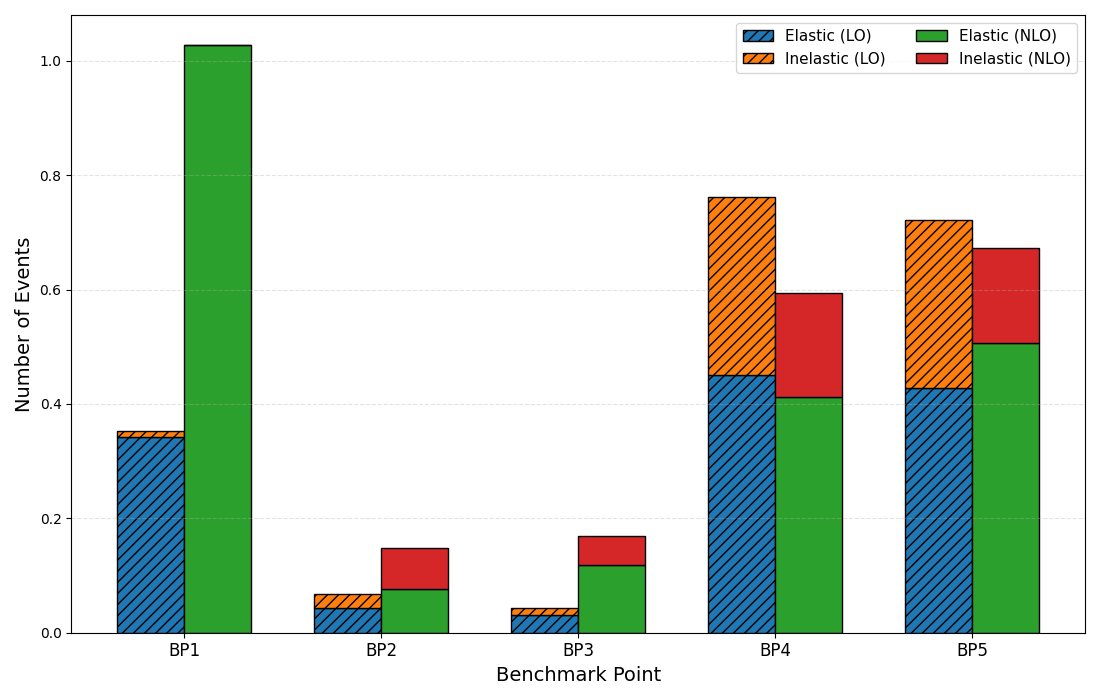}
    \caption{The figure illustrates the total number of events for each benchmark point.
    The contributions from the elastic and inelastic scattering
    are shown in separate colors. BP1-BP5 refer to the benchmarks discussed in Table
    \ref{tab:bp1}. The left bar (blue-orange) for each BP shows the uncorrected LO
    contribution and the right bar (green-red) shows the NLO contribution.
    }
    \label{fig:rate}
\end{figure}

Finally, we evaluate the number of scattering events in the detector material by summing the contributions from both elastic and inelastic processes. For this purpose we use the recently reported exposure of the LZ experiment ($M_Dt_{\rm ex}=4.2$\,tonne\,$\cdot$\,yr) \cite{LZCollaboration:2024lux}. The results are listed in Table~\ref{tab:rate}, where we present the predicted event rates at both LO and NLO precision.
For comparison, Fig.~\ref{fig:rate} displays these results as a bar chart for the benchmark scenarios. The blue (orange) bars represent the number of inelastic events at LO (NLO), while the green (red) bars correspond to the elastic events at LO (NLO) as mentioned in the legend.
{Our implementation of the LZ DD constraints reproduces the null result of the
experiment \cite{LZCollaboration:2024lux}.}

\begin{table}[h!]
	\centering
    \renewcommand{\arraystretch}{1.2}
	\caption{Number of events corresponding to elastic and inelastic scattering, evaluated at
    LO and NLO, along with the predicted total events in the detector. The required exposure
    ($\mathcal E=M_Dt_{\rm ex}$) to probe the BPs is also listed.} 
	\begin{tabular}{lcccccccc}
    \hline
    Benchmark & BP1 & BP2 & BP3 & BP4 & BP5 \\
    \hline
    Elastic events (LO)
    & {0.3419} & 0.0436 & 0.0312 & 0.4506 & 0.4272 \\
    
    Inelastic events (LO)
    & {0.0108} & 0.02471 & 0.0112 & 0.31205 & 0.2947 \\
    \hline
    
    Elastic events (NLO)
    & {1.0280} & 0.07729 & 0.1176 & 0.4127 & 0.5074 \\
    
    Inelastic events (NLO)
    & {0.00022} & 0.07039 & 0.0518 & 0.1806 & 0.1661 \\
    
    Combined events (NLO)
    & {1.02822} & 0.14768 & 0.1694 & 0.5933 & 0.6735 \\
    
    
    Required Exposure ($M_Dt_{\rm ex}$) [t-y]
     & {9.41} & 65.50 & 57.10 & 16.30 & 14.36 \\

    
    \hline
    \end{tabular}
	\label{tab:rate}
\end{table}

For BP1, the elastic rate dominates the total event yield due to its significantly
larger cross-section. In contrast, due to a {relatively larger inelastic
cross-section of BP2, the contributions from inelastic scattering approach 100\% of
the contributions from elastic scattering events. Even though
$\sigma_{\rm NLO}^{\rm el} < \sigma_{\rm NLO}^{\rm inel}$, but still the event
contributions from the inelastic channel is comparatively a little less due to the
Boltzmann suppression arising from the increased $v_{\rm min}^{\rm inel}$ in the
inelastic channel.
} For BPs 3-5, the elastic
scattering provides larger contributions
due to their larger cross-section. Additionally, the inelastic event rate is
Boltzmann suppressed due to the requirement of a minimum DM kinetic energy (or
equivalently, velocity) to excite the heavier state. As a result, the inelastic
event rate remains small compared to the elastic contribution. Nevertheless, for
these benchmark points, the inelastic channel provides a non-negligible contribution
to the total number of scattering events.


\section{Summary and Conclusion}
\label{sec:Conclusion}


In this work, a comprehensive study of the spin-independent direct detection of Higgsino-like neutralino DM in the Minimal Supersymmetric Standard Model has been presented, incorporating both elastic and inelastic scattering channels. While conventional direct-detection analyses typically focus on elastic scattering, the presence of a nearly degenerate Higgsino-like next-to-lightest neutralino can render inelastic processes phenomenologically relevant when the mass splitting between the two lightest neutralinos is comparable to the typical kinetic energy of galactic DM particles.

To obtain precise predictions for the direct-detection rates, we have computed the complete set of one-loop electroweak corrections to the ($\tilde{\chi}_1^0\tilde{\chi}_i^0 h_j$) vertices ($i,j\in{1,2})$, including the corresponding counterterm contributions within an on-shell renormalization framework. Since the inelastic scattering rate is highly sensitive to the mass splitting ($\delta = m_{\tilde{\chi}_2^0}-m_{\tilde{\chi}_1^0}$), we have consistently evaluated the one-loop corrected neutralino and chargino masses using an appropriate on-shell renormalization scheme. The resulting on-shell masses and couplings have subsequently been implemented within \texttt{micrOMEGAs}, extending its standard treatment of direct detection to include radiatively corrected inelastic Higgs-mediated processes.

Our analysis demonstrates that radiative corrections can substantially modify both the neutralino mass splitting and the effective Higgs couplings, thereby significantly affecting the relative importance of elastic and inelastic scattering channels. In particular, for scenarios with mass splittings of ($\mathcal{O}(100)\,\mathrm{keV}$) or smaller, the
{inelastic contribution to scattering events in the detector can become comparable
to the corresponding elastic contribution (BP2).} Consequently, a consistent
interpretation of direct-detection limits and future sensitivities in compressed
Higgsino scenarios requires the simultaneous consideration of both channels.

The results presented here highlight the importance of precision calculations for Higgsino-like $\tilde{\chi}_1^0$ DM in the era of present and next-generation direct-detection experiments. The DM DD experiments at present and in the near future will provide significantly enhanced exposures and improved sensitivities to spin-independent DM--nucleon cross-sections. These experiments are expected to probe cross-sections approaching the neutrino floor over a broad range of DM masses, including the TeV-scale region relevant for Higgsino-like DM.

Given their large target masses and increased exposure, future experiments will also become increasingly sensitive to compressed electroweakino spectra in which inelastic up-scattering is kinematically accessible. 
{The recently reported LZ nuclear-recoil event near 248 keV provides an additional motivation for studying this possibility. While its interpretation as a dark-matter signal remains premature, recent studies have pointed out that TeV-scale Higgsino dark matter can be relevant in this context \cite{Wu:2026nhi, Freese:2026sga, Fan:2026kxx}. A dedicated analysis of this event within the radiatively corrected inelastic Higgsino framework developed here, including the recoil-energy spectrum and the combined elastic and inelastic contributions, would therefore be an interesting direction for future work.}
Consequently, radiatively corrected predictions for both elastic and inelastic scattering processes will be essential for robust interpretations of experimental results and for accurately assessing the discovery prospects of Higgsino-like $\tilde{\chi}_1^0$ DM. Beyond enhancing the sensitivity of direct-detection searches, these studies provide a complementary avenue to probe natural supersymmetric scenarios, which are characterized by a compressed Higgsino spectrum and suppressed tree-level interactions. Therefore, future direct-detection experiments, { such as XENONnT, PandaX, and LZ future runs,} combined with
precision theoretical predictions, will not only extend the reach for DM discovery but also provide a complementary probe of parameter regions favored by the natural SUSY paradigm that
are difficult to access through other experimental avenues.

\section*{Acknowledgments}
{The work of AP was conducted under the state assignment of Lomonosov State University.}
The authors thank Prof. Genevieve Bélanger for the useful discussions.
The authors acknowledge Dr. Subhadip Bisal for his contributions towards the early stages
of the work. The computations in this project were partially supported by SAMKHYA,
the high-performance computing (HPC) facility provided by the Institute of Physics,
Bhubaneswar (IOPB).

\bigskip
\appendix
\section*{Appendix}
\section{Kinematics}
\label{Kinematics}
Let the 3-momentum of the incoming dark matter particle in the lab frame be $\vec{p}_1$,
of mass $m_{\chi_1}$, and let its energy be $E_1$. Let the mass of the target nucleus be
$m_A$, which is at rest initially in the lab frame, and its energy is $E_A$. Let the
3-momentum of $\chi_2$, produced after collision be $\vec{p}_1'$, with energy $E_1'$ in
the lab frame. The 3-momentum of the nucleus after scattering is given by $\vec{q}$, and
the energy of the nucleus after collision is $E_A'$ in the lab frame. Let $\theta$ be the
recoil angle in lab frame, and $\delta=m_{\chi_2}-m_{\chi_1}$ is the mass splitting of the
two neutralino states $\chi_1$ and $\chi_2$.

Momentum transfer in the lab frame
\begin{align}
	\vec{q} = \vec{p}_1-\vec{p}_1' \text{ and }
	q^2=2 m_A E_R
\end{align}
where $E_R$ is the recoil energy.
Energy conservation in the non-relativistic (NR) limit in the lab frame yields
\begin{align}
	\frac{p_1^2}{2m_{\chi_1}} = \delta+\frac{p_1'^2}
	{2m_{\chi_2}}+\frac{q^2}{2m_A}.
\end{align}
The minimum velocity of the DM to produce a recoil inelastically is
\begin{align}
	\implies (v_{\chi_1})_{\rm min}(E_R)&=  \frac{1}{q}\left(\frac{q^2}{2\mu_A}+\delta\right) \\
	\implies (v_{\chi_1})_{\rm min}(E_R)&=  {v_{\rm min}(E_R)=\frac{1}{\sqrt{2m_AE_R}}
		\left(\frac{E_R m_A}{\mu_A}+\delta\right)} \label{eq:vmin_app}
\end{align}
putting $q=\sqrt{2m_NE_R}$. {The function $v_{\rm min}(E_R)$ is bounded from below by the kinematic limit of velocity, ${\rm Min}(v_{\rm min}(E_R))=\sqrt{2\delta/\mu_A}$ at $E_R=\mu_A\delta/m_A$. 
Additionally, the requirement $v_{\rm min}(E_R)<v_{\rm max}$ sets limits on the maximum and minimum recoil energy, for which the Heaviside $\Theta$ function in Eq.~\eqref{1} is satisfied. These limits can be evaluated as,
\begin{align}\label{eq:E_R}
		(E_R)_{\text{max}} =
		\frac{\mu_A^2}{2m_A}
		\left(v_{\rm max} + \sqrt{v_{\rm max}^2 - \frac{2\delta}{\mu_A}}\right)^2
	,
	\qquad
		(E_R)_{\text{min}} =
		\frac{\mu_A^2}{2m_A}
		\left(v_{\rm max} - \sqrt{v_{\rm max}^2 - \frac{2\delta}{\mu_A}}\right)^2
	.
\end{align}

While in this work, the integration order is first over $v$, then $E_R$, one might adopt a reverse order of integration such that,
\begin{align}
    \int_{E_{R}^{\rm min}}^{E_{R}^{\rm max}}dE_R\int_{v_{\rm min}(E_R)}^{v_{\rm max}}dv\dots~
    \to~
    \int_{v_{\rm min}}^{v_{\rm max}} dv\int_{E_{R}^{\rm min}(v)}^{E_{R}^{\rm max}(v)}dE_R\dots
\end{align}
where the limits in the new order are given as,
\begin{align}
v_{\rm min}&=\sqrt{\frac{2\delta}{\mu_A}}\\
	(E_R)_{\text{max}}(v) =
		\frac{\mu_A^2}{2m_A}
		\left(v + \sqrt{v^2 - \frac{2\delta}{\mu_A}}\right)^2
	&,
	\qquad
		(E_R)_{\text{min}}(v) =
		\frac{\mu_A^2}{2m_A}
		\left(v - \sqrt{v^2 - \frac{2\delta}{\mu_A}}\right)^2
	.
    \end{align}
    The same results as those presented in this work are obtained using
    this approach, and this has been explicitly cross-checked.}

\section{Direct Detection} \label{sec:velocities}
The Maxwell-Boltzmann distribution function for the velocity of Dark Matter, keeping into
account the Earth's velocity ($v_e$) in the galaxy, is given as:
\begin{align}\label{3}
	f(v) = \frac{v}{v_e v_0 \sqrt{\pi}} \left[e^{-(v-v_e)^2/v_0^2}-e^{-(v+v_e)^2/v_0^2}
	\right].
\end{align}
{$f(v)$ is normalized as
\begin{align}
    \int_0^{\infty} f(v) dv = 1.
\end{align}

In our calculations, we have taken the upper limit of the DM velocity integral
to be infinity for simplicity. This approximation may lead to a slight
overestimation of the DM flux reaching the detector for large $\delta$.
However, for the $\mathcal{O}(10)$~keV mass splittings considered
in this work and for a xenon-based LZ detector, this effect is
negligible, as discussed in Ref.~\cite{Tucker-Smith:2001myb}.
}

$v_e$ is made up of two components since the Sun moves around the center of the galaxy,
while the Earth moves around the Sun:
\begin{align}
	v_e = v_0 \left[1.05 + 0.07 \cos\left(\frac{2 \pi (t - t_p)}{1 \mathrm{yr}}\right)
		\right]
\end{align}

We define dimensionless variables $\eta=v_e/v_0$ and $x_{\rm min}=v_{\rm min}/v_0$.
$t_p \simeq 2^{\rm nd}$ June, and yr denotes year.
Inserting Eqs.~(\ref{2}) and (\ref{3}) in Eq.~(\ref{1}), and performing the velocity
integration will fetch one \cite{Tucker-Smith:2001myb}{
\begin{align}\label{eq:rate}
	\frac{dR_{\rm SI}}{dE_R} = \frac{N_T m_A \rho_{\odot}}{4 v_0 m_{\tilde{\chi}_1^0}} F^2(E_R)
	\frac{\sigma_{p}}{\mu_{p}^2}\frac{(f_p Z+f_n(A-Z))^2}{f_p^2}
	\left(\frac{\mathrm{erf}(x_{\rm min}+\eta)-\mathrm{erf}(x_{\rm min}-\eta)}{\eta}\right),
\end{align}
{where $Z$ and $A$ are the atomic number and mass number of the target nuclei, respectively
($Z=54, ~A=131$ for Xe used in LZ).}

The differential rate depends on the mass splitting $\delta$ via $x_{\rm min}$,
given by
\begin{align}\label{eq:xmin}
	x_{\rm min} = & \frac{v_{\rm min}}{v_0} = \frac{1}{v_0} \sqrt{\frac{1}{2 m_A E_R}} \left(
	 \frac{m_A E_R}{\mu_A} + \delta \right),  
\end{align}
where $\mu_A$} is the reduced mass of the relic-nucleus system. {In the limit 
$\delta \rightarrow 0$, the expressions for $v_{\rm min}$ and the differential rate in the
inelastic scenario reduce to those of the elastic case.} A non-zero $\delta$
increases the minimum DM particle speed required to produce a given nuclear recoil energy.
In Eq.~(\ref{eq:rate}), the {DM-nucleon scattering} cross-section is given as{
\begin{align}
	\sigma_{p} = \frac{4}{\pi} \, 
	\mu_{p}^2 \, |f_{p}|^2\label{eq:InelCS}
\end{align}
where $f_{p}$ is the amplitude of the inelastic proton scattering process 
and $\mu_{p}$ is the relic-proton reduced mass.}

Therefore, in the case of inelastic scattering, for a given mass, the minimum velocity
required for the dark matter to scatter off a nucleus and get excited to a slightly
heavier state needs to be increased to obtain the same recoil energy as the elastic
scattering one. This is due to the fact that the outgoing dark matter state has more
mass (and hence more momentum), which needs to be made up for by the scattering velocity.


\bigskip
\bibliographystyle{JHEPCust.bst}
\bibliography{Higgsino}

@article{Das:2020ozo,
    author = "Das, Debottam and De, Bibhabasu and Mitra, Subhadip",
    title = "{Cancellation in Dark Matter-Nucleon Interactions: the Role of Non-Standard-Model-like Yukawa Couplings}",
    eprint = "2011.13225",
    archivePrefix = "arXiv",
    primaryClass = "hep-ph",
    reportNumber = "IP-BBSR/2020-6",
    doi = "10.1016/j.physletb.2021.136159",
    journal = "Phys. Lett. B",
    volume = "815",
    pages = "136159",
    year = "2021"
}

@article{Cheung:2013dua,
    author = "Cheung, Clifford and Sanford, David",
    title = "{Simplified Models of Mixed Dark Matter}",
    eprint = "1311.5896",
    archivePrefix = "arXiv",
    primaryClass = "hep-ph",
    reportNumber = "CALT-68-2870",
    doi = "10.1088/1475-7516/2014/02/011",
    journal = "JCAP",
    volume = "02",
    pages = "011",
    year = "2014"
}

@article{Huang:2014xua,
    author = "Huang, Peisi and Wagner, Carlos E. M.",
    title = "{Blind Spots for neutralino Dark Matter in the MSSM with an intermediate $m_A$}",
    eprint = "1404.0392",
    archivePrefix = "arXiv",
    primaryClass = "hep-ph",
    doi = "10.1103/PhysRevD.90.015018",
    journal = "Phys. Rev. D",
    volume = "90",
    number = "1",
    pages = "015018",
    year = "2014"
}

@article{Cheung:2012qy,
    author = "Cheung, Clifford and Hall, Lawrence J. and Pinner, David and Ruderman, Joshua T.",
    title = "{Prospects and Blind Spots for Neutralino Dark Matter}",
    eprint = "1211.4873",
    archivePrefix = "arXiv",
    primaryClass = "hep-ph",
    doi = "10.1007/JHEP05(2013)100",
    journal = "JHEP",
    volume = "05",
    pages = "100",
    year = "2013"
}

@article{Cohen:2013ama,
    author = "Cohen, Timothy and Lisanti, Mariangela and Pierce, Aaron and Slatyer, Tracy R.",
    title = "{Wino Dark Matter Under Siege}",
    eprint = "1307.4082",
    archivePrefix = "arXiv",
    primaryClass = "hep-ph",
    reportNumber = "MIT-CTP-4482, SLAC-PUB-15664, MCTP-13-19",
    doi = "10.1088/1475-7516/2013/10/061",
    journal = "JCAP",
    volume = "10",
    pages = "061",
    year = "2013"
}

@article{Hisano:2010ct,
    author = "Hisano, Junji and Ishiwata, Koji and Nagata, Natsumi",
    title = "{Gluon contribution to the dark matter direct detection}",
    eprint = "1007.2601",
    archivePrefix = "arXiv",
    primaryClass = "hep-ph",
    reportNumber = "IPMU10-0113, ICRR-REPORT-570-2010-3",
    doi = "10.1103/PhysRevD.82.115007",
    journal = "Phys. Rev. D",
    volume = "82",
    pages = "115007",
    year = "2010"
}

@article{Fan:2013faa,
    author = "Fan, JiJi and Reece, Matthew",
    title = "{In Wino Veritas? Indirect Searches Shed Light on Neutralino Dark Matter}",
    eprint = "1307.4400",
    archivePrefix = "arXiv",
    primaryClass = "hep-ph",
    doi = "10.1007/JHEP10(2013)124",
    journal = "JHEP",
    volume = "10",
    pages = "124",
    year = "2013"
}

@article{MAGIC:2022acl,
    author = "Abe, H. and others",
    collaboration = "MAGIC",
    title = "{Search for Gamma-Ray Spectral Lines from Dark Matter Annihilation up to 100~TeV toward the Galactic Center with MAGIC}",
    eprint = "2212.10527",
    archivePrefix = "arXiv",
    primaryClass = "astro-ph.HE",
    reportNumber = "KEK-TH-2487, KEK-Cosmo-0307",
    doi = "10.1103/PhysRevLett.130.061002",
    journal = "Phys. Rev. Lett.",
    volume = "130",
    number = "6",
    pages = "061002",
    year = "2023"
}

@article{Chattopadhyay:2003xi,
    author = "Chattopadhyay, Utpal and Corsetti, Achille and Nath, Pran",
    title = "{WMAP constraints, SUSY dark matter and implications for the direct detection of SUSY}",
    eprint = "hep-ph/0303201",
    archivePrefix = "arXiv",
    doi = "10.1103/PhysRevD.68.035005",
    journal = "Phys. Rev. D",
    volume = "68",
    pages = "035005",
    year = "2003"
}

@article{Ellis:2012aa,
    author = "Ellis, John and Olive, Keith A.",
    title = "{Revisiting the Higgs Mass and Dark Matter in the CMSSM}",
    eprint = "1202.3262",
    archivePrefix = "arXiv",
    primaryClass = "hep-ph",
    reportNumber = "KCL-PH-TH-2012-08, LCTS-2012-04, CERN-PH-TH-2012-045, UMN-TH-3034-12, FTPI-MINN-12-07, UMN--TH--3034-12, FTPI--MINN--12-07",
    doi = "10.1140/epjc/s10052-012-2005-2",
    journal = "Eur. Phys. J. C",
    volume = "72",
    pages = "2005",
    year = "2012"
}

@article{Baer:2011ab,
    author = "Baer, Howard and Barger, Vernon and Mustafayev, Azar",
    title = "{Implications of a 125 GeV Higgs scalar for LHC SUSY and neutralino dark matter searches}",
    eprint = "1112.3017",
    archivePrefix = "arXiv",
    primaryClass = "hep-ph",
    reportNumber = "UMN-TH-3024-11, FTPI-MINN-11-32",
    doi = "10.1103/PhysRevD.85.075010",
    journal = "Phys. Rev. D",
    volume = "85",
    pages = "075010",
    year = "2012"
}

@article{Akula:2011jx,
    author = "Akula, Sujeet and Liu, Mengxi and Nath, Pran and Peim, Gregory",
    title = "{Naturalness, Supersymmetry and Implications for LHC and Dark Matter}",
    eprint = "1111.4589",
    archivePrefix = "arXiv",
    primaryClass = "hep-ph",
    doi = "10.1016/j.physletb.2012.01.077",
    journal = "Phys. Lett. B",
    volume = "709",
    pages = "192--199",
    year = "2012"
}

@article{Chan:1997bi,
    author = "Chan, Kwok Lung and Chattopadhyay, Utpal and Nath, Pran",
    title = "{Naturalness, weak scale supersymmetry and the prospect for the observation of supersymmetry at the Tevatron and at the CERN LHC}",
    eprint = "hep-ph/9710473",
    archivePrefix = "arXiv",
    doi = "10.1103/PhysRevD.58.096004",
    journal = "Phys. Rev. D",
    volume = "58",
    pages = "096004",
    year = "1998"
}

@article{He:2023lgi,
    author = "He, Yangle and Meng, Lei and Yue, Yuanfang and Zhang, Di",
    title = "{Impact of the recent measurement of (g-2){\ensuremath{\mu}}, the LHC search for supersymmetry, and the LZ experiment on the minimal supersymmetric standard model}",
    eprint = "2303.02360",
    archivePrefix = "arXiv",
    primaryClass = "hep-ph",
    doi = "10.1103/PhysRevD.108.115010",
    journal = "Phys. Rev. D",
    volume = "108",
    number = "11",
    pages = "115010",
    year = "2023"
}

@article{Chattopadhyay:2005mv,
    author = "Chattopadhyay, Utpal and Choudhury, Debajyoti and Drees, Manuel and Konar, Partha and Roy, D. P.",
    title = "{Looking for a heavy Higgsino LSP in collider and dark matter experiments}",
    eprint = "hep-ph/0508098",
    archivePrefix = "arXiv",
    reportNumber = "TIFR-TH-05-13",
    doi = "10.1016/j.physletb.2005.09.088",
    journal = "Phys. Lett. B",
    volume = "632",
    pages = "114--126",
    year = "2006"
}

@article{Buchmueller:2013rsa,
    author = "Buchmueller, O. and others",
    title = "{The CMSSM and NUHM1 after LHC Run 1}",
    eprint = "1312.5250",
    archivePrefix = "arXiv",
    primaryClass = "hep-ph",
    reportNumber = "KCL-PH-TH-2013-42, LCTS-2013-29, CERN-PH-TH-2013-297, DESY-13-250, UMN-TH-3316-13, SLAC-PUB-15861, FERMILAB-PUB-13-668-T, FTPI-MINN-13-43",
    doi = "10.1140/epjc/s10052-014-2922-3",
    journal = "Eur. Phys. J. C",
    volume = "74",
    number = "6",
    pages = "2922",
    year = "2014"
}

@article{WMAP:2012nax,
    author = "Hinshaw, G. and others",
    collaboration = "WMAP",
    title = "{Nine-Year Wilkinson Microwave Anisotropy Probe (WMAP) Observations: Cosmological Parameter Results}",
    eprint = "1212.5226",
    archivePrefix = "arXiv",
    primaryClass = "astro-ph.CO",
    doi = "10.1088/0067-0049/208/2/19",
    journal = "Astrophys. J. Suppl.",
    volume = "208",
    pages = "19",
    year = "2013"
}

@article{Planck:2018vyg,
    author = "Aghanim, N. and others",
    collaboration = "Planck",
    title = "{Planck 2018 results. VI. Cosmological parameters}",
    eprint = "1807.06209",
    archivePrefix = "arXiv",
    primaryClass = "astro-ph.CO",
    doi = "10.1051/0004-6361/201833910",
    journal = "Astron. Astrophys.",
    volume = "641",
    pages = "A6",
    year = "2020",
    note = "[Erratum: Astron.Astrophys. 652, C4 (2021)]"
}

@article{Hall:1997ah,
    author = "Hall, Lawrence J. and Moroi, Takeo and Murayama, Hitoshi",
    title = "{Sneutrino cold dark matter with lepton number violation}",
    eprint = "hep-ph/9712515",
    archivePrefix = "arXiv",
    reportNumber = "LBNL-41199, LBL-41199, UCB-PTH-97-69",
    doi = "10.1016/S0370-2693(98)00196-8",
    journal = "Phys. Lett. B",
    volume = "424",
    pages = "305--312",
    year = "1998"
}

@article{Alguero:2021dig,
  author        = {Alguero, Ga\"el and Heisig, Jan and Khosa, Charanjit and Kraml, Sabine and Kulkarni, Suchita and Lessa, Andre and Reyes-Gonz\'alez, Humberto and Waltenberger, Wolfgang and Wongel, Alicia},
  title         = {{Constraining new physics with SModelS version 2}},
  eprint        = {2112.00769},
  archiveprefix = {arXiv},
  primaryclass  = {hep-ph},
  reportnumber  = {TTK-21-50},
  month         = {12},
  year          = {2021}
}

@article{Aparicio:2016qqb,
  author        = {Aparicio, Luis and Cicoli, Michele and Dutta, Bhaskar and Muia, Francesco and Quevedo, Fernando},
  title         = {{Light Higgsino Dark Matter from Non-thermal Cosmology}},
  eprint        = {1607.00004},
  archiveprefix = {arXiv},
  primaryclass  = {hep-ph},
  doi           = {10.1007/JHEP11(2016)038},
  journal       = {JHEP},
  volume        = {11},
  pages         = {038},
  year          = {2016}
}

@article{Bae:2013hma,
  author        = {Bae, Kyu Jung and Baer, Howard and Chun, Eung Jin},
  title         = {{Mixed axion/neutralino dark matter in the SUSY DFSZ axion model}},
  eprint        = {1309.5365},
  archiveprefix = {arXiv},
  primaryclass  = {hep-ph},
  doi           = {10.1088/1475-7516/2013/12/028},
  journal       = {JCAP},
  volume        = {12},
  pages         = {028},
  year          = {2013}
}

@article{Baer:2012up,
    author = "Baer, Howard and Barger, Vernon and Huang, Peisi and Mustafayev, Azar and Tata, Xerxes",
    title = "{Radiative natural SUSY with a 125 GeV Higgs boson}",
    eprint = "1207.3343",
    archivePrefix = "arXiv",
    primaryClass = "hep-ph",
    doi = "10.1103/PhysRevLett.109.161802",
    journal = "Phys. Rev. Lett.",
    volume = "109",
    pages = "161802",
    year = "2012"
}

@article{Baer:2012cf,
  author        = {Baer, Howard and Barger, Vernon and Huang, Peisi and Mickelson, Dan and Mustafayev, Azar and Tata, Xerxes},
  title         = {{Radiative natural supersymmetry: Reconciling electroweak fine-tuning and the Higgs boson mass}},
  eprint        = {1212.2655},
  archiveprefix = {arXiv},
  primaryclass  = {hep-ph},
  doi           = {10.1103/PhysRevD.87.115028},
  journal       = {Phys. Rev. D},
  volume        = {87},
  number        = {11},
  pages         = {115028},
  year          = {2013}
}

@article{Kerr:1986cei,
    author = "Kerr, F. J. and Lynden-Bell, D.",
    title = "{Review of galactic constants}",
    doi = "10.1093/mnras/221.4.1023",
    journal = "Mon. Not. Roy. Astron. Soc.",
    volume = "221",
    number = "4",
    pages = "1023--1038",
    year = "1986"
}

@article{Barbieri:1987fn,
  author       = {Barbieri, Riccardo and Giudice, G. F.},
  title        = {{Upper Bounds on Supersymmetric Particle Masses}},
  journal      = {Nucl. Phys.},
  volume       = {B306},
  year         = {1988},
  pages        = {63-76},
  doi          = {10.1016/0550-3213(88)90171-X},
  reportnumber = {CERN-TH-4825/87},
  slaccitation = {%%CITATION = NUPHA,B306,63;%%}
}

@article{Baro:2007em,
  author        = {Baro, N. and Boudjema, F. and Semenov, A.},
  title         = {{Full one-loop corrections to the relic density in the MSSM: A Few examples}},
  eprint        = {0710.1821},
  archiveprefix = {arXiv},
  primaryclass  = {hep-ph},
  reportnumber  = {LAPTH-1211-07},
  doi           = {10.1016/j.physletb.2008.01.031},
  journal       = {Phys. Lett. B},
  volume        = {660},
  pages         = {550--560},
  year          = {2008}
}

@article{Baro:2009na,
  author        = {Baro, N. and Boudjema, F. and Chalons, G. and Hao, Sun},
  title         = {{Relic density at one-loop with gauge boson pair production}},
  eprint        = {0910.3293},
  archiveprefix = {arXiv},
  primaryclass  = {hep-ph},
  doi           = {10.1103/PhysRevD.81.015005},
  journal       = {Phys. Rev. D},
  volume        = {81},
  pages         = {015005},
  year          = {2010}
}

@article{Baro2009,
  title     = {Automatized full one-loop renormalization of the MSSM. II. The chargino-neutralino sector, the sfermion sector, and some applications},
  author    = {Baro, N. and Boudjema, F.},
  journal   = {Phys. Rev. D},
  volume    = {80},
  issue     = {7},
  pages     = {076010},
  numpages  = {24},
  year      = {2009},
  month     = {Oct},
  publisher = {American Physical Society},
  doi       = {10.1103/PhysRevD.80.076010},
  url       = {https://link.aps.org/doi/10.1103/PhysRevD.80.076010}
}

@article{Belanger:2004yn,
  author        = {Belanger, G. and Boudjema, F. and Pukhov, A. and Semenov, A.},
  title         = {{micrOMEGAs: Version 1.3}},
  eprint        = {hep-ph/0405253},
  archiveprefix = {arXiv},
  reportnumber  = {LAPTH-1044},
  doi           = {10.1016/j.cpc.2005.12.005},
  journal       = {Comput. Phys. Commun.},
  volume        = {174},
  pages         = {577--604},
  year          = {2006}
}

@article{Belanger:2006is,
  author        = {Belanger, G. and Boudjema, F. and Pukhov, A. and Semenov, A.},
  title         = {{MicrOMEGAs 2.0: A Program to calculate the relic density of dark matter in a generic model}},
  eprint        = {hep-ph/0607059},
  archiveprefix = {arXiv},
  reportnumber  = {LAPTH-1152-06},
  doi           = {10.1016/j.cpc.2006.11.008},
  journal       = {Comput. Phys. Commun.},
  volume        = {176},
  pages         = {367--382},
  year          = {2007}
}

@article{Belanger:2008sj,
  author        = {Belanger, G. and Boudjema, F. and Pukhov, A. and Semenov, A.},
  title         = {{Dark matter direct detection rate in a generic model with micrOMEGAs 2.2}},
  eprint        = {0803.2360},
  archiveprefix = {arXiv},
  primaryclass  = {hep-ph},
  reportnumber  = {LAPTH-1237-08},
  doi           = {10.1016/j.cpc.2008.11.019},
  journal       = {Comput. Phys. Commun.},
  volume        = {180},
  pages         = {747--767},
  year          = {2009}
}

@article{Belanger:2010pz,
  author        = {Belanger, G. and Boudjema, F. and Pukhov, A. and Semenov, A.},
  title         = {{micrOMEGAs: A Tool for dark matter studies}},
  eprint        = {1005.4133},
  archiveprefix = {arXiv},
  primaryclass  = {hep-ph},
  doi           = {10.1393/ncc/i2010-10591-3},
  journal       = {Nuovo Cim. C},
  volume        = {033N2},
  pages         = {111--116},
  year          = {2010}
}

@article{Belanger:2013oya,
  author        = {Belanger, G. and Boudjema, F. and Pukhov, A. and Semenov, A.},
  title         = {{micrOMEGAs$\_$3: A program for calculating dark matter observables}},
  eprint        = {1305.0237},
  archiveprefix = {arXiv},
  primaryclass  = {hep-ph},
  reportnumber  = {LAPTH-023-13},
  doi           = {10.1016/j.cpc.2013.10.016},
  journal       = {Comput. Phys. Commun.},
  volume        = {185},
  pages         = {960--985},
  year          = {2014}
}

@article{Bertone:2004pz,
  author        = {Bertone, Gianfranco and Hooper, Dan and Silk, Joseph},
  title         = {{Particle dark matter: Evidence, candidates and constraints}},
  eprint        = {hep-ph/0404175},
  archiveprefix = {arXiv},
  reportnumber  = {FERMILAB-PUB-04-047-A},
  doi           = {10.1016/j.physrep.2004.08.031},
  journal       = {Phys. Rept.},
  volume        = {405},
  pages         = {279--390},
  year          = {2005}
}

@article{Bisal:2023fgb,
  author        = {Bisal, Subhadip and Chatterjee, Arindam and Das, Debottam and Pasha, Syed Adil},
  title         = {{Radiative corrections to aid the direct detection of the Higgsino-like neutralino dark matter: Spin-independent interactions}},
  eprint        = {2311.09937},
  archiveprefix = {arXiv},
  primaryclass  = {hep-ph},
  doi           = {10.1103/PhysRevD.110.023043},
  journal       = {Phys. Rev. D},
  volume        = {110},
  number        = {2},
  pages         = {023043},
  year          = {2024}
}

@article{Bisal:2023iip,
    author = "Bisal, Subhadip and Chatterjee, Arindam and Das, Debottam and Pasha, Syed Adil",
    title = "{Confronting electroweak MSSM through one-loop renormalized neutralino-Higgs interactions for dark matter direct detection and the muon g-2}",
    eprint = "2311.09938",
    archivePrefix = "arXiv",
    primaryClass = "hep-ph",
    doi = "10.1103/PhysRevD.110.015021",
    journal = "Phys. Rev. D",
    volume = "110",
    number = "1",
    pages = "015021",
    year = "2024"
}

@article{Bisal:2024ezn,
  author        = {Bisal, Subhadip and Chatterjee, Arindam and Das, Debottam and Pasha, Syed Adil},
  title         = {{Radiative corrections to the direct detection of the Higgsino- and wino-like neutralino dark matter: Spin-dependent interactions}},
  eprint        = {2410.18205},
  archiveprefix = {arXiv},
  primaryclass  = {hep-ph},
  doi           = {10.1103/PhysRevD.111.083003},
  journal       = {Phys. Rev. D},
  volume        = {111},
  number        = {8},
  pages         = {083003},
  year          = {2025}
}

@article{Bisal:2024iar,
  author        = {Bisal, Subhadip and Chatterjee, Arindam and Das, Debottam and Pasha, Syed Adil},
  title         = {{Electroweak renormalization of neutralino-Higgs interactions at one loop and its impacts on spin-independent direct detection of Wino-like dark matter}},
  eprint        = {2410.18206},
  archiveprefix = {arXiv},
  primaryclass  = {hep-ph},
  doi           = {10.1103/PhysRevD.111.055021},
  journal       = {Phys. Rev. D},
  volume        = {111},
  number        = {5},
  pages         = {055021},
  year          = {2025}
}

@article{Chakraborti:2015mra,
  author        = {Chakraborti, Manimala and Chattopadhyay, Utpal and Choudhury, Arghya and Datta, Amitava and Poddar, Sujoy},
  title         = {{Reduced LHC constraints for higgsino-like heavier electroweakinos}},
  eprint        = {1507.01395},
  archiveprefix = {arXiv},
  primaryclass  = {hep-ph},
  reportnumber  = {HRI-RECAPP-2015-014},
  doi           = {10.1007/JHEP11(2015)050},
  journal       = {JHEP},
  volume        = {11},
  pages         = {050},
  year          = {2015}
}

@article{Chatterjee:2011wc,
  author        = {Chatterjee, Arindam and Drees, Manuel and Kulkarni, Suchita and Xu, Qingjun},
  title         = {{On the On-Shell Renormalization of the Chargino and Neutralino Masses in the MSSM}},
  eprint        = {1107.5218},
  archiveprefix = {arXiv},
  primaryclass  = {hep-ph},
  doi           = {10.1103/PhysRevD.85.075013},
  journal       = {Phys. Rev. D},
  volume        = {85},
  pages         = {075013},
  year          = {2012}
}

@article{Chatterjee:2012hkk,
  author        = {Chatterjee, Arindam and Drees, Manuel and Kulkarni, Suchita},
  title         = {{Radiative Corrections to the Neutralino Dark Matter Relic Density - an Effective Coupling Approach}},
  eprint        = {1209.2328},
  archiveprefix = {arXiv},
  primaryclass  = {hep-ph},
  doi           = {10.1103/PhysRevD.86.105025},
  journal       = {Phys. Rev. D},
  volume        = {86},
  pages         = {105025},
  year          = {2012}
}

@article{Cirelli:2005uq,
  author        = {Cirelli, Marco and Fornengo, Nicolao and Strumia, Alessandro},
  title         = {{Minimal dark matter}},
  eprint        = {hep-ph/0512090},
  archiveprefix = {arXiv},
  reportnumber  = {DFTT40-2005, IFUP-TH-2005-34},
  doi           = {10.1016/j.nuclphysb.2006.07.012},
  journal       = {Nucl. Phys. B},
  volume        = {753},
  pages         = {178--194},
  year          = {2006}
}

@article{Chattopadhyay:2024qgs,
    author = "Chattopadhyay, Utpal and Das, Debottam and Poddar, Sujoy and Puri, Rahul and Saha, Abhijit Kumar",
    title = "{Implications of Sgr A$^{*}$ on the \ensuremath{\gamma}-rays searches of Bino dark matter with (g-2)$_{\ensuremath{\mu}}$}",
    eprint = "2407.14603",
    archivePrefix = "arXiv",
    primaryClass = "hep-ph",
    doi = "10.1088/1475-7516/2025/01/121",
    journal = "JCAP",
    volume = "01",
    pages = "121",
    year = "2025"
}

@article{Hisano:2006nn,
    author = "Hisano, Junji and Matsumoto, Shigeki and Nagai, Minoru and Saito, Osamu and Senami, Masato",
    title = "{Non-perturbative effect on thermal relic abundance of dark matter}",
    eprint = "hep-ph/0610249",
    archivePrefix = "arXiv",
    reportNumber = "KEK-TH-1111",
    doi = "10.1016/j.physletb.2007.01.012",
    journal = "Phys. Lett. B",
    volume = "646",
    pages = "34--38",
    year = "2007"
}

@article{Beneke:2014hja,
    author = "Beneke, M. and Hellmann, Charlotte and Ruiz-Femenia, P.",
    title = "{Heavy neutralino relic abundance with Sommerfeld enhancements - a study of pMSSM scenarios}",
    eprint = "1411.6930",
    archivePrefix = "arXiv",
    primaryClass = "hep-ph",
    reportNumber = "TUM-HEP-955-14, TTK-14-22, SFB-CPP-14-70, IFIC-14-59",
    doi = "10.1007/JHEP03(2015)162",
    journal = "JHEP",
    volume = "03",
    pages = "162",
    year = "2015"
}

@article{Hisano:2004ds,
    author = "Hisano, Junji and Matsumoto, Shigeki. and Nojiri, Mihoko M. and Saito, Osamu",
    title = "{Non-perturbative effect on dark matter annihilation and gamma ray signature from galactic center}",
    eprint = "hep-ph/0412403",
    archivePrefix = "arXiv",
    reportNumber = "ICRR-REPORT-513-2004-11, YITP-04-73",
    doi = "10.1103/PhysRevD.71.063528",
    journal = "Phys. Rev. D",
    volume = "71",
    pages = "063528",
    year = "2005"
}

@article{Mohanty:2010es,
    author = "Mohanty, Subhendra and Rao, Soumya and Roy, D. P.",
    title = "{Relic density and PAMELA events in a heavy wino dark matter model with Sommerfeld effect}",
    eprint = "1009.5058",
    archivePrefix = "arXiv",
    primaryClass = "hep-ph",
    doi = "10.1142/S0217751X1250025X",
    journal = "Int. J. Mod. Phys. A",
    volume = "27",
    number = "6",
    pages = "1250025",
    year = "2012"
}

@article{Beneke:2019qaa,
    author = "Beneke, Martin and Szafron, Robert and Urban, Kai",
    title = "{Wino potential and Sommerfeld effect at NLO}",
    eprint = "1909.04584",
    archivePrefix = "arXiv",
    primaryClass = "hep-ph",
    reportNumber = "TUM-HEP-1222/19",
    doi = "10.1016/j.physletb.2019.135112",
    journal = "Phys. Lett. B",
    volume = "800",
    pages = "135112",
    year = "2020"
}

@article{Fritzsche:2011nr,
    author = "Fritzsche, T. and Heinemeyer, S. and Rzehak, H. and Schappacher, C.",
    title = "{Heavy Scalar Top Quark Decays in the Complex MSSM: A Full One-Loop Analysis}",
    eprint = "1111.7289",
    archivePrefix = "arXiv",
    primaryClass = "hep-ph",
    reportNumber = "CERN-PH-TH-2011-291, FR-PHENO-2011-021, KA-TP-34-2011",
    doi = "10.1103/PhysRevD.86.035014",
    journal = "Phys. Rev. D",
    volume = "86",
    pages = "035014",
    year = "2012"
}

@article{Bisal:2026hpm,
    author = "Bisal, Subhadip and Chatterjee, Arindam and Das, Debottam and Pasha, Syed Adil and Puri, Rahul",
    title = "{Unveiling the Vanishing Higgsino-Nucleon Scattering in the MSSM at Next-to-Leading Order}",
    eprint = "2607.20588",
    archivePrefix = "arXiv",
    primaryClass = "hep-ph",
    month = "7",
    year = "2026"
}

@article{Fan:2026kxx,
    author = "Fan, JiJi and Reece, Matthew",
    title = "{Higgsino Above the Sea of Fog}",
    eprint = "2609.01504",
    archivePrefix = "arXiv",
    primaryClass = "hep-ph",
    month = "9",
    year = "2026"
}

@article{Freese:2026sga,
    author = "Freese, Katherine and Theodosopoulos, Dionysios P.",
    title = "{Higgsino Dark Matter Interpretation of the LUX-ZEPLIN 248 keV Nuclear-Recoil Event}",
    eprint = "2609.01583",
    archivePrefix = "arXiv",
    primaryClass = "hep-ph",
    month = "9",
    year = "2026"
}

@article{Wu:2026nhi,
    author = "Wu, Lei and Zhang, Yang and Zhu, Bin",
    title = "{TeV Higgsino Dark Matter from LZ Nuclear Recoil to Fermi-LAT Gamma Rays}",
    eprint = "2609.01590",
    archivePrefix = "arXiv",
    primaryClass = "hep-ph",
    month = "9",
    year = "2026"
}

@article{Akerib:2026jyz,
    author = "Akerib, D. S. and others",
    title = "{Search for dark matter particle interactions in an extended nuclear recoil energy window with the LUX-ZEPLIN (LZ) experiment}",
    eprint = "2609.02823",
    archivePrefix = "arXiv",
    primaryClass = "hep-ex",
    month = "9",
    year = "2026"
}

@article{Djouadi:2000ck,
  author        = {Djouadi, Abdelhak and Drees, Manuel},
  title         = {{QCD corrections to neutralino nucleon scattering}},
  eprint        = {hep-ph/0004205},
  archiveprefix = {arXiv},
  reportnumber  = {PM-00-14, TUM-HEP-370-00},
  doi           = {10.1016/S0370-2693(00)00661-4},
  journal       = {Phys. Lett. B},
  volume        = {484},
  pages         = {183--191},
  year          = {2000}
}

@article{Drees:1996pk,
  author        = {Drees, Manuel and Nojiri, Mihoko M. and Roy, D. P. and Yamada, Youichi},
  title         = {{Light Higgsino dark matter}},
  eprint        = {hep-ph/9701219},
  archiveprefix = {arXiv},
  reportnumber  = {APCTP-96-06, KEK-TH-505, KEK-PREPRINT-96-156, TIFR-TH-96-62, TU-515},
  doi           = {10.1103/PhysRevD.64.039901},
  journal       = {Phys. Rev. D},
  volume        = {56},
  pages         = {276--290},
  year          = {1997},
  note          = {[Erratum: Phys.Rev.D 64, 039901 (2001)]}
}

@book{Drees:2004jm,
    author = {Drees, Manuel and Godbole, Rohini and Roy, Probir},
    title = {Theory and Phenomenology of Sparticles},
    publisher = {WORLD SCIENTIFIC},
    year = {2005},
    doi = {10.1142/4001},
    address = {},
    edition   = {},
    URL = {https://www.worldscientific.com/doi/abs/10.1142/4001},
    eprint = {https://www.worldscientific.com/doi/pdf/10.1142/4001}
}

@article{Drees:1993bu,
    author = "Drees, Manuel and Nojiri, Mihoko",
    title = "{Neutralino - nucleon scattering revisited}",
    eprint = "hep-ph/9307208",
    archivePrefix = "arXiv",
    reportNumber = "MAD-PH-768",
    doi = "10.1103/PhysRevD.48.3483",
    journal = "Phys. Rev. D",
    volume = "48",
    pages = "3483--3501",
    year = "1993"
}

@article{ATLAS:2022pib,
    author = "Aad, Georges and others",
    collaboration = "ATLAS",
    title = "{Search for heavy, long-lived, charged particles with large ionisation energy loss in $pp$ collisions at $\sqrt{s} = 13~\text{TeV}$ using the ATLAS experiment and the full Run 2 dataset}",
    eprint = "2205.06013",
    archivePrefix = "arXiv",
    primaryClass = "hep-ex",
    reportNumber = "CERN-EP-2022-029",
    doi = "10.1007/JHEP06(2023)158",
    journal = "JHEP",
    volume = "2306",
    pages = "158",
    year = "2023"
}

@article{Drees1,
  title     = {Neutralino-nucleon scattering reexamined},
  author    = {Drees, Manuel and Nojiri, Mihoko M.},
  journal   = {Phys. Rev. D},
  volume    = {48},
  issue     = {8},
  pages     = {3483--3501},
  numpages  = {0},
  year      = {1993},
  month     = {Oct},
  publisher = {American Physical Society},
  doi       = {10.1103/PhysRevD.48.3483},
  url       = {https://link.aps.org/doi/10.1103/PhysRevD.48.3483}
}

@article{Drees2,
  title     = {New contributions to coherent neutralino-nucleus scattering},
  author    = {Drees, Manuel and Nojiri, Mihoko M.},
  journal   = {Phys. Rev. D},
  volume    = {47},
  issue     = {10},
  pages     = {4226--4232},
  numpages  = {0},
  year      = {1993},
  month     = {May},
  publisher = {American Physical Society},
  doi       = {10.1103/PhysRevD.47.4226},
  url       = {https://link.aps.org/doi/10.1103/PhysRevD.47.4226}
}

@article{Ellis:1986yg,
  author  = {Ellis, J and Enqvist, K. and Nanopoulos, D.V. and Zwirner, F.},
  title   = {Observables in   low-energy superstring models},
  journal = {Modern Physics Letters A},
  volume  = {01},
  number  = {01},
  pages   = {57-69},
  year    = {1986},
  doi     = {10.1142/S0217732386000105},
  url     = {https://doi.org/10.1142/S0217732386000105},
  eprint  = {https://doi.org/10.1142/S0217732386000105}
}

@article{ELLIS1991,
  author  = {John Ellis and Ricardo A. Flores},
  title   = {Elastic supersymmetric relic-nucleus scattering revisited},
  journal = {Physics Letters B},
  volume  = {263},
  number  = {2},
  pages   = {259-266},
  year    = {1991},
  issn    = {0370-2693},
  doi     = {https://doi.org/10.1016/0370-2693(91)90597-J},
  url     = {https://www.sciencedirect.com/science/article/pii/037026939190597J}
}

@article{ELLIS1993,
  title   = {Prospects for neutralino detection with a 73Ge + 76Ge detector},
  journal = {Physics Letters B},
  volume  = {300},
  number  = {1},
  pages   = {175-182},
  year    = {1993},
  issn    = {0370-2693},
  doi     = {https://doi.org/10.1016/0370-2693(93)90767-C},
  url     = {https://www.sciencedirect.com/science/article/pii/037026939390767C},
  author  = {John Ellis and Ricardo A. Flores}
}

@article{Feng:2013pwa,
  author        = {Feng, Jonathan L.},
  title         = {{Naturalness and the Status of Supersymmetry}},
  journal       = {Ann. Rev. Nucl. Part. Sci.},
  volume        = {63},
  year          = {2013},
  pages         = {351-382},
  doi           = {10.1146/annurev-nucl-102010-130447},
  eprint        = {1302.6587},
  archiveprefix = {arXiv},
  primaryclass  = {hep-ph},
  reportnumber  = {UCI-TR-2013-01},
  slaccitation  = {%%CITATION = ARXIV:1302.6587;%%}
}

@article{Fritzsche:2002bi,
  author        = {Fritzsche, T. and Hollik, W.},
  title         = {{Complete one loop corrections to the mass spectrum of charginos and neutralinos in the MSSM}},
  eprint        = {hep-ph/0203159},
  archiveprefix = {arXiv},
  reportnumber  = {KA-TP-6-2002},
  doi           = {10.1007/s10052-002-0992-0},
  journal       = {Eur. Phys. J. C},
  volume        = {24},
  pages         = {619--629},
  year          = {2002}
}

@article{Fritzsche:2013fta,
  author        = {Fritzsche, T. and Hahn, T. and Heinemeyer, S. and von der Pahlen, F. and Rzehak, H. and Schappacher, C.},
  title         = {{The Implementation of the Renormalized Complex MSSM in FeynArts and FormCalc}},
  eprint        = {1309.1692},
  archiveprefix = {arXiv},
  primaryclass  = {hep-ph},
  reportnumber  = {CERN-PH-TH-2013-168, MPP-2013-185},
  doi           = {10.1016/j.cpc.2014.02.005},
  journal       = {Comput. Phys. Commun.},
  volume        = {185},
  pages         = {1529--1545},
  year          = {2014}
}

@article{Giudice:2013nak,
  author        = {Giudice, Gian F.},
  title         = {{Naturalness after LHC8}},
  booktitle     = {{Proceedings, 2013 European Physical Society Conference
                   on High Energy Physics (EPS-HEP 2013): Stockholm, Sweden,
                   July 18-24, 2013}},
  journal       = {PoS},
  volume        = {EPS-HEP2013},
  year          = {2013},
  pages         = {163},
  eprint        = {1307.7879},
  archiveprefix = {arXiv},
  primaryclass  = {hep-ph},
  reportnumber  = {CERN-PH-TH-2013-180},
  slaccitation  = {%%CITATION = ARXIV:1307.7879;%%}
}

@article{HAHN1999153,
  title    = {Automated one-loop calculations in four and D dimensions},
  journal  = {Computer Physics Communications},
  volume   = {118},
  number   = {2},
  pages    = {153-165},
  year     = {1999},
  issn     = {0010-4655},
  doi      = {https://doi.org/10.1016/S0010-4655(98)00173-8},
  url      = {https://www.sciencedirect.com/science/article/pii/S0010465598001738},
  author   = {T. Hahn and M. Pérez-Victoria}
}

@article{HAHN2001418,
  title    = {Generating Feynman diagrams and amplitudes with FeynArts 3},
  journal  = {Computer Physics Communications},
  volume   = {140},
  number   = {3},
  pages    = {418-431},
  year     = {2001},
  issn     = {0010-4655},
  doi      = {https://doi.org/10.1016/S0010-4655(01)00290-9},
  url      = {https://www.sciencedirect.com/science/article/pii/S0010465501002909},
  author   = {Thomas Hahn}
}

@article{Harz:2014gaa,
  author        = {Harz, J. and Herrmann, B. and Klasen, M. and Kova\v{r}\'\i{}k, K. and Meinecke, M.},
  title         = {{SUSY-QCD corrections to stop annihilation into electroweak final states including Coulomb enhancement effects}},
  eprint        = {1410.8063},
  archiveprefix = {arXiv},
  primaryclass  = {hep-ph},
  doi           = {10.1103/PhysRevD.91.034012},
  journal       = {Phys. Rev. D},
  volume        = {91},
  number        = {3},
  pages         = {034012},
  year          = {2015}
}

@article{Harz:2014tma,
  author        = {Harz, J. and Herrmann, B. and Klasen, M. and Kovarik, K.},
  title         = {{One-loop corrections to neutralino-stop coannihilation revisited}},
  eprint        = {1409.2898},
  archiveprefix = {arXiv},
  primaryclass  = {hep-ph},
  reportnumber  = {LAPTH-049-14, LCTS-2014-33, MS-TP-14-24},
  doi           = {10.1103/PhysRevD.91.034028},
  journal       = {Phys. Rev. D},
  volume        = {91},
  number        = {3},
  pages         = {034028},
  year          = {2015}
}

@article{Harz:2023llw,
  author        = {Harz, Julia and Herrmann, Bj\"orn and Klasen, Michael and Kova\v{r}\'\i{}k, Karol and Wiggering, Luca Paolo},
  title         = {{Precision predictions for dark matter with DM@NLO in the MSSM}},
  eprint        = {2312.17206},
  archiveprefix = {arXiv},
  primaryclass  = {hep-ph},
  reportnumber  = {MITP-23-084, MS-TP-23-52, LAPTH-062/23},
  month         = {12},
  year          = {2023}
}

@article{Hisano:2004pv,
  author        = {Hisano, Junji and Matsumoto, Shigeki and Nojiri, Mihoko M. and Saito, Osamu},
  title         = {{Direct detection of the Wino and Higgsino-like neutralino dark matters at one-loop level}},
  eprint        = {hep-ph/0407168},
  archiveprefix = {arXiv},
  reportnumber  = {ICRR-REPORT-506-2004-4, YITP-04-39},
  doi           = {10.1103/PhysRevD.71.015007},
  journal       = {Phys. Rev. D},
  volume        = {71},
  pages         = {015007},
  year          = {2005}
}

@article{Hisano:2010fy,
  author        = {Hisano, Junji and Ishiwata, Koji and Nagata, Natsumi},
  title         = {{A complete calculation for direct detection of Wino dark matter}},
  eprint        = {1004.4090},
  archiveprefix = {arXiv},
  primaryclass  = {hep-ph},
  reportnumber  = {IPMU-10-0066, ICRR-REPORT-568-2010-1},
  doi           = {10.1016/j.physletb.2010.05.047},
  journal       = {Phys. Lett. B},
  volume        = {690},
  pages         = {311--315},
  year          = {2010}
}

@article{Hisano:2011cs,
  author        = {Hisano, Junji and Ishiwata, Koji and Nagata, Natsumi and Takesako, Tomohiro},
  title         = {{Direct Detection of Electroweak-Interacting Dark Matter}},
  eprint        = {1104.0228},
  archiveprefix = {arXiv},
  primaryclass  = {hep-ph},
  reportnumber  = {IPMU-11-0046, ICRR-REPORT-583-2010-16, CALT-68-2824},
  doi           = {10.1007/JHEP07(2011)005},
  journal       = {JHEP},
  volume        = {07},
  pages         = {005},
  year          = {2011}
}

@article{Hisano:2012wm,
  author        = {Hisano, Junji and Ishiwata, Koji and Nagata, Natsumi},
  title         = {{Direct Search of Dark Matter in High-Scale Supersymmetry}},
  eprint        = {1210.5985},
  archiveprefix = {arXiv},
  primaryclass  = {hep-ph},
  reportnumber  = {IPMU-12-0187, CALT-68-2891},
  doi           = {10.1103/PhysRevD.87.035020},
  journal       = {Phys. Rev. D},
  volume        = {87},
  pages         = {035020},
  year          = {2013}
}

@article{Jungman:1995df,
  author        = {Jungman, Gerard and Kamionkowski, Marc and Griest, Kim},
  title         = {{Supersymmetric dark matter}},
  eprint        = {hep-ph/9506380},
  archiveprefix = {arXiv},
  reportnumber  = {SU-4240-605, UCSD-PTH-95-02, IASSNS-HEP-95-14, CU-TP-677},
  doi           = {10.1016/0370-1573(95)00058-5},
  journal       = {Phys. Rept.},
  volume        = {267},
  pages         = {195--373},
  year          = {1996}
}

@article{Klasen:2016qyz,
  author        = {Klasen, Michael and Kovarik, Karol and Steppeler, Patrick},
  title         = {{SUSY-QCD corrections for direct detection of neutralino dark matter and correlations with relic density}},
  eprint        = {1607.06396},
  archiveprefix = {arXiv},
  primaryclass  = {hep-ph},
  reportnumber  = {MS-TP-16-14},
  doi           = {10.1103/PhysRevD.94.095002},
  journal       = {Phys. Rev. D},
  volume        = {94},
  number        = {9},
  pages         = {095002},
  year          = {2016}
}

@article{Kraml:2013mwa,
  author        = {Kraml, Sabine and Kulkarni, Suchita and Laa, Ursula and
                   Lessa, Andre and Magerl, Wolfgang and Proschofsky, Doris and
                   Waltenberger, Wolfgang},
  title         = {{SModelS: a tool for interpreting simplified-model
                   results from the LHC and its application to
                   supersymmetry}},
  journal       = {Eur.Phys.J.},
  volume        = {C74},
  pages         = {2868},
  doi           = {10.1140/epjc/s10052-014-2868-5},
  year          = {2014},
  eprint        = {1312.4175},
  archiveprefix = {arXiv},
  primaryclass  = {hep-ph},
  slaccitation  = {%%CITATION = ARXIV:1312.4175;%%}
}

@article{KUBLBECK1990165,
  title    = {Feyn arts — computer-algebraic generation of Feynman graphs and amplitudes},
  journal  = {Computer Physics Communications},
  volume   = {60},
  number   = {2},
  pages    = {165-180},
  year     = {1990},
  issn     = {0010-4655},
  doi      = {https://doi.org/10.1016/0010-4655(90)90001-H},
  url      = {https://www.sciencedirect.com/science/article/pii/001046559090001H},
  author   = {J. Küblbeck and M. Böhm and A. Denner}
}

@article{LEWIN199687,
  title    = {Review of mathematics, numerical factors, and corrections for dark matter experiments based on elastic nuclear recoil},
  journal  = {Astroparticle Physics},
  volume   = {6},
  number   = {1},
  pages    = {87-112},
  year     = {1996},
  issn     = {0927-6505},
  doi      = {https://doi.org/10.1016/S0927-6505(96)00047-3},
  url      = {https://www.sciencedirect.com/science/article/pii/S0927650596000473},
  author   = {J.D. Lewin and P.F. Smith}
}

@article{LZCollaboration:2024lux,
  author        = {Aalbers, J. and others},
  collaboration = {LZ Collaboration},
  title         = {{Dark Matter Search Results from 4.2 Tonne-Years of Exposure of the LUX-ZEPLIN (LZ) Experiment}},
  eprint        = {2410.17036},
  journal       = {Phys. Rev. Lett.},
  volume        = {135},
  issue         = {1},
  pages         = {011802},
  numpages      = {11},
  year          = {2025},
  month         = {Jul},
  publisher     = {American Physical Society},
  doi           = {10.1103/4dyc-z8zf},
  url           = {https://link.aps.org/doi/10.1103/4dyc-z8zf}
}

@article{Mustafayev:2014lqa,
  author        = {Mustafayev, Azar and Tata, Xerxes},
  title         = {{Supersymmetry, Naturalness, and Light Higgsinos}},
  journal       = {Indian J. Phys.},
  volume        = {88},
  year          = {2014},
  pages         = {991-1004},
  doi           = {10.1007/s12648-014-0504-8},
  eprint        = {1404.1386},
  archiveprefix = {arXiv},
  primaryclass  = {hep-ph},
  reportnumber  = {UH-511-1231-14},
  slaccitation  = {%%CITATION = ARXIV:1404.1386;%%}
}

@article{PandaX,
  title         = {{Search for Light Dark Matter with Ionization Signals in the PandaX-4T Experiment}},
  author        = {{Li, Shuaijie and Wu, Mengmeng and Abdukerim, Abdusalam and Bo, Zihao and Chen, Wei and Chen, Xun and Chen, Yunhua and Cheng, Chen and Cheng, Zhaokan and Cui, Xiangyi and Fan, Yingjie and Fang, Deqing and Fu, Changbo and Fu, Mengting and Geng, Lisheng and Giboni, Karl and Gu, Linhui and Guo, Xuyuan and Han, Chencheng and Han, Ke and He, Changda and He, Jinrong and Huang, Di and Huang, Yanlin and Huang, Zhou and Hou, Ruquan and Ji, Xiangdong and Ju, Yonglin and Li, Chenxiang and Li, Jiafu and Li, Mingchuan and Li, Shu and Lin, Qing and Liu, Jianglai and Lu, Xiaoying and Luo, Lingyin and Luo, Yunyang and Ma, Wenbo and Ma, Yugang and Mao, Yajun and Meng, Yue and Ning, Xuyang and Qi, Ningchun and Qian, Zhicheng and Ren, Xiangxiang and Shaheed, Nasir and Shang, Changsong and Shang, Xiaofeng and Shen, Guofang and Si, Lin and Sun, Wenliang and Tan, Andi and Tao, Yi and Wang, Anqing and Wang, Meng and Wang, Qiuhong and Wang, Shaobo and Wang, Siguang and Wang, Wei and Wang, Xiuli and Wang, Zhou and Wei, Yuehuan and Wu, Weihao and Xia, Jingkai and Xiao, Mengjiao and Xiao, Xiang and Xie, Pengwei and Yan, Binbin and Yan, Xiyu and Yang, Jijun and Yang, Yong and Yao, Yukun and You, Zhengyun and Yu, Chunxu and Yuan, Jumin and Yuan, Ying and Yuan, Zhe and Zeng, Xinning and Zhang, Dan and Zhang, Minzhen and Zhang, Peng and Zhang, Shibo and Zhang, Shu and Zhang, Tao and Zhang, Yang and Zhang, Yingxin and Zhang, Yuanyuan and Zhao, Li and Zheng, Qibin and Zhou, Jifang and Zhou, Ning and Zhou, Xiaopeng and Zhou, Yong and Zhou, Yubo}},
  collaboration = {PandaX Collaboration},
  journal       = {Phys. Rev. Lett.},
  volume        = {130},
  issue         = {26},
  pages         = {261001},
  numpages      = {7},
  year          = {2023},
  month         = {Jun},
  publisher     = {American Physical Society},
  doi           = {10.1103/PhysRevLett.130.261001},
  url           = {https://link.aps.org/doi/10.1103/PhysRevLett.130.261001}
}

@article{PASSARINO1979151,
  title   = {One-loop corrections for e+e- annihilation into $\mu^+ \mu^-$ in the Weinberg model},
  journal = {Nuclear Physics B},
  volume  = {160},
  number  = {1},
  pages   = {151-207},
  year    = {1979},
  issn    = {0550-3213},
  doi     = {https://doi.org/10.1016/0550-3213(79)90234-7},
  url     = {https://www.sciencedirect.com/science/article/pii/0550321379902347},
  author  = {G. Passarino and M. Veltman}
}

@article{Drees:2006um,
    author = "Drees, Manuel and Hollik, Wolfgang and Xu, Qingjun",
    title = "{One-loop calculations of the decay of the next-to-lightest neutralino in the MSSM}",
    eprint = "hep-ph/0610267",
    archivePrefix = "arXiv",
    reportNumber = "MPP-2006-117, FREIBURG-THEP-06-16",
    doi = "10.1088/1126-6708/2007/02/032",
    journal = "JHEP",
    volume = "02",
    pages = "032",
    year = "2007"
}

@article{Heinemeyer:2011gk,
    author = "Heinemeyer, S. and von der Pahlen, F. and Schappacher, C.",
    title = "{Chargino Decays in the Complex MSSM: A Full One-Loop Analysis}",
    eprint = "1112.0760",
    archivePrefix = "arXiv",
    primaryClass = "hep-ph",
    reportNumber = "KA-TP-38-2011",
    doi = "10.1140/epjc/s10052-012-1892-6",
    journal = "Eur. Phys. J. C",
    volume = "72",
    pages = "1892",
    year = "2012"
}

@article{Bharucha:2012re,
    author = "Bharucha, A. and Heinemeyer, S. and von der Pahlen, F. and Schappacher, C.",
    title = "{Neutralino Decays in the Complex MSSM at One-Loop: a Comparison of On-Shell Renormalization Schemes}",
    eprint = "1208.4106",
    archivePrefix = "arXiv",
    primaryClass = "hep-ph",
    doi = "10.1103/PhysRevD.86.075023",
    journal = "Phys. Rev. D",
    volume = "86",
    pages = "075023",
    year = "2012"
}

@article{Oller:2003ge,
    author = "Oller, W. and Eberl, H. and Majerotto, W. and Weber, C.",
    title = "{Analysis of the chargino and neutralino mass parameters at one loop level}",
    eprint = "hep-ph/0304006",
    archivePrefix = "arXiv",
    reportNumber = "HEPHY-PUB-769-03",
    doi = "10.1140/epjc/s2003-01246-9",
    journal = "Eur. Phys. J. C",
    volume = "29",
    pages = "563--572",
    year = "2003"
}

@article{Oller:2005xg,
    author = "Oller, W. and Eberl, H. and Majerotto, W.",
    title = "{Precise predictions for chargino and neutralino pair production in e+ e- annihilation}",
    eprint = "hep-ph/0504109",
    archivePrefix = "arXiv",
    reportNumber = "HEPHY-PUB-806-05",
    doi = "10.1103/PhysRevD.71.115002",
    journal = "Phys. Rev. D",
    volume = "71",
    pages = "115002",
    year = "2005"
}

@article{Eberl:2001iw,
  author       = {Eberl, H. and Kincel, M. and Majerotto, W. and Yamada, Y.},
  title        = {One loop corrections to the chargino and neutralino mass matrices in the on-shell scheme},
  journal      = {Phys. Rev. D},
  volume       = {64},
  pages        = {115013},
  year         = {2001},
  eprint       = {hep-ph/0104109},
  archivePrefix= {arXiv},
  doi          = {10.1103/PhysRevD.64.115013}
}

@article{Bahl:2022igd,
    author = {Bahl, Henning and Biek{\"o}tter, Thomas and Heinemeyer, Sven and Li, Cheng and Paasch, Steven and Weiglein, Georg and Wittbrodt, Jonas},
    title = "{HiggsTools: BSM scalar phenomenology with new versions of HiggsBounds and HiggsSignals}",
    eprint = "2210.09332",
    archivePrefix = "arXiv",
    primaryClass = "hep-ph",
    doi = "10.1016/j.cpc.2023.108803",
    journal = "Comput. Phys. Commun.",
    volume = "291",
    pages = "108803",
    year = "2023"
}

@article{Bechtle:2020pkv,
    author = "Bechtle, Philip and Dercks, Daniel and Heinemeyer, Sven and Klingl, Tobias and Stefaniak, Tim and Weiglein, Georg and Wittbrodt, Jonas",
    title = "{HiggsBounds-5: Testing Higgs Sectors in the LHC 13 TeV Era}",
    eprint = "2006.06007",
    archivePrefix = "arXiv",
    primaryClass = "hep-ph",
    reportNumber = "BONN-TH-2020-03, DESY 20-093, DESY-20-093, IFT-UAM/CSIC-20-072, LU 20-27",
    doi = "10.1140/epjc/s10052-020-08557-9",
    journal = "Eur. Phys. J. C",
    volume = "80",
    number = "12",
    pages = "1211",
    year = "2020"
}

@article{Bechtle:2020uwn,
    author = "Bechtle, Philip and Heinemeyer, Sven and Klingl, Tobias and Stefaniak, Tim and Weiglein, Georg and Wittbrodt, Jonas",
    title = "{HiggsSignals-2: Probing new physics with precision Higgs measurements in the LHC 13 TeV era}",
    eprint = "2012.09197",
    archivePrefix = "arXiv",
    primaryClass = "hep-ph",
    reportNumber = "BONN-TH-2020-09, DESY-20-228, DESY 20-228, IFT-UAM/CSIC-20-081, LU TP 20-53",
    doi = "10.1140/epjc/s10052-021-08942-y",
    journal = "Eur. Phys. J. C",
    volume = "81",
    number = "2",
    pages = "145",
    year = "2021"
}

@article{Porod:2003um,
  author        = {Porod, Werner},
  title         = {{SPheno, a program for calculating supersymmetric spectra, SUSY particle decays and SUSY particle production at e+ e- colliders}},
  eprint        = {hep-ph/0301101},
  archiveprefix = {arXiv},
  reportnumber  = {ZU-TH-01-03},
  doi           = {10.1016/S0010-4655(03)00222-4},
  journal       = {Comput. Phys. Commun.},
  volume        = {153},
  pages         = {275--315},
  year          = {2003}
}

@article{Staub:2013tta,
  author        = {Staub, Florian},
  title         = {{SARAH 4 : A tool for (not only SUSY) model builders}},
  eprint        = {1309.7223},
  archiveprefix = {arXiv},
  primaryclass  = {hep-ph},
  reportnumber  = {BONN-TH-2013-17},
  doi           = {10.1016/j.cpc.2014.02.018},
  journal       = {Comput. Phys. Commun.},
  volume        = {185},
  pages         = {1773--1790},
  year          = {2014}
}

@article{Staub:2015kfa,
  author        = {Staub, Florian},
  title         = {{Exploring new models in all detail with SARAH}},
  eprint        = {1503.04200},
  archiveprefix = {arXiv},
  primaryclass  = {hep-ph},
  reportnumber  = {CERN-PH-TH-2015-051},
  doi           = {10.1155/2015/840780},
  journal       = {Adv. High Energy Phys.},
  volume        = {2015},
  pages         = {840780},
  year          = {2015}
}

@article{Staub:2017jnp,
  author        = {Staub, Florian and Porod, Werner},
  title         = {{Improved predictions for intermediate and heavy Supersymmetry in the MSSM and beyond}},
  eprint        = {1703.03267},
  archiveprefix = {arXiv},
  primaryclass  = {hep-ph},
  reportnumber  = {KA-TP-03-2017},
  doi           = {10.1140/epjc/s10052-017-4893-7},
  journal       = {Eur. Phys. J. C},
  volume        = {77},
  number        = {5},
  pages         = {338},
  year          = {2017}
}

@article{STAUB20141773,
  title    = {SARAH   4: A tool for (not only SUSY) model builders},
  journal  = {Computer Physics Communications},
  volume   = {185},
  number   = {6},
  pages    = {1773-1790},
  year     = {2014},
  issn     = {0010-4655},
  doi      = {https://doi.org/10.1016/j.cpc.2014.02.018},
  url      = {https://www.sciencedirect.com/science/article/pii/S0010465514000629},
  author   = {Florian Staub}
}

@article{Tucker-Smith:2001myb,
  author        = {Tucker-Smith, David and Weiner, Neal},
  title         = {{Inelastic dark matter}},
  eprint        = {hep-ph/0101138},
  archiveprefix = {arXiv},
  reportnumber  = {UCB-PTH-00-43, LBNL-47234, UW-PT-00-17},
  doi           = {10.1103/PhysRevD.64.043502},
  journal       = {Phys. Rev. D},
  volume        = {64},
  pages         = {043502},
  year          = {2001}
}

@article{Altakach:2024jwk,
    author = "Altakach, Mohammad Mahdi and Kraml, Sabine and Lessa, Andre and Narasimha, Sahana and Pascal, Timoth{\'e}e and Ramos, Camila and Villamizar, Yoxara and Waltenberger, Wolfgang",
    title = "{SModelS v3: going beyond $ \mathcal{Z} _{2}$ topologies}",
    eprint = "2409.12942",
    archivePrefix = "arXiv",
    primaryClass = "hep-ph",
    doi = "10.1007/JHEP11(2024)074",
    journal = "JHEP",
    volume = "11",
    pages = "074",
    year = "2024"
}

@article{MahdiAltakach:2023bdn,
    author = "Mahdi Altakach, Mohammad and Kraml, Sabine and Lessa, Andre and Narasimha, Sahana and Pascal, Timoth{\'e}e and Waltenberger, Wolfgang",
    title = "{SModelS v2.3: Enabling global likelihood analyses}",
    eprint = "2306.17676",
    archivePrefix = "arXiv",
    primaryClass = "hep-ph",
    doi = "10.21468/SciPostPhys.15.5.185",
    journal = "SciPost Phys.",
    volume = "15",
    number = "5",
    pages = "185",
    year = "2023"
}

@article{Tucker-Smith:2004mxa,
    author = "Tucker-Smith, David and Weiner, Neal",
    title = "{The Status of inelastic dark matter}",
    eprint = "hep-ph/0402065",
    archivePrefix = "arXiv",
    reportNumber = "UW-PT-04-01",
    doi = "10.1103/PhysRevD.72.063509",
    journal = "Phys. Rev. D",
    volume = "72",
    pages = "063509",
    year = "2005"
}

@article{Dessert:2022evk,
    author = "Dessert, Christopher and Foster, Joshua W. and Park, Yujin and Safdi, Benjamin R. and Xu, Weishuang Linda",
    title = "{Higgsino Dark Matter Confronts 14~Years of Fermi {\ensuremath{\gamma}}-Ray Data}",
    eprint = "2207.10090",
    archivePrefix = "arXiv",
    primaryClass = "hep-ph",
    reportNumber = "MIT-CTP/5454",
    doi = "10.1103/PhysRevLett.130.201001",
    journal = "Phys. Rev. Lett.",
    volume = "130",
    number = "20",
    pages = "201001",
    year = "2023"
}

@article{ATLAS:2025lhc,
    author = "Aad, Georges and others",
    collaboration = "ATLAS",
    title = "{Search for higgsinos in compressed mass spectra using low-momentum tracks in $pp$ collisions at $\sqrt{s}=13$ TeV with the ATLAS detector}",
    eprint = "2511.20042",
    archivePrefix = "arXiv",
    primaryClass = "hep-ex",
    reportNumber = "CERN-EP-2025-246",
    month = "11",
    year = "2025"
}

@article{Engel:1991wq,
    author = "Engel, J.",
    title = "{Nuclear form-factors for the scattering of weakly interacting massive particles}",
    doi = "10.1016/0370-2693(91)90712-Y",
    journal = "Phys. Lett. B",
    volume = "264",
    pages = "114--119",
    year = "1991"
}

\end{document}